%% file: cvpr.tex
\documentclass[final]{cvpr}

\usepackage{times}
\usepackage{epsfig}
\usepackage{graphicx}
\usepackage{amsmath}
\usepackage{amssymb}

\input{vcl-shortcuts.tex}
\usepackage[pagebackref=true,breaklinks=true,colorlinks,bookmarks=false]{hyperref}

\def\cvprPaperID{3736} 
\def\confYear{CVPR 2021}

\begin{document}

\title{Invertible Image Signal Processing}

\author{{Yazhou Xing\thanks{Joint first authors} 
\quad \quad
Zian Qian\footnotemark[1] 

\quad \quad
Qifeng Chen}\\
The Hong Kong University of Science and Technology\\
}

\maketitle

\thispagestyle{empty}
\pagestyle{empty}

\begin{abstract}

Unprocessed RAW data is a highly valuable image format for image editing and computer vision. However, since the file size of RAW data is huge, most users can only get access to processed and compressed sRGB images. To bridge this gap, we design an Invertible Image Signal Processing (InvISP) pipeline, which not only enables rendering visually appealing sRGB images but also allows recovering nearly perfect RAW data. Due to our framework's inherent reversibility, we can reconstruct realistic RAW data instead of synthesizing RAW data from sRGB images without any memory overhead. We also integrate a differentiable JPEG compression simulator that empowers our framework to reconstruct RAW data from JPEG images. Extensive quantitative and qualitative experiments on two DSLR demonstrate that our method obtains much higher quality in both rendered sRGB images and reconstructed RAW data than alternative methods. 
\end{abstract}


\begin{figure*}[t!]
\centering
\includegraphics[width=1\textwidth]{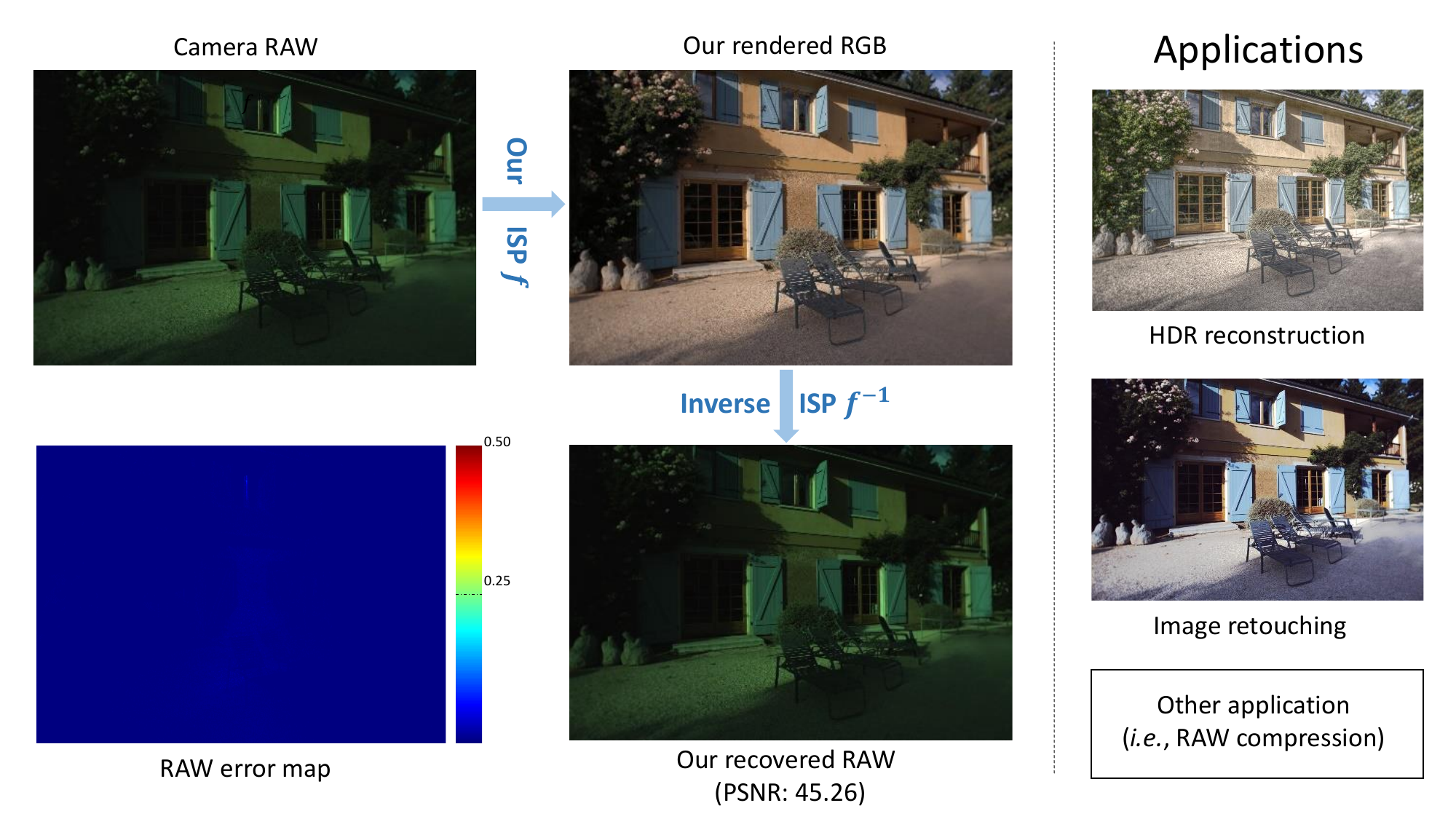}
\vspace{-3mm}
\caption{Our ISP model can not only render visually pleasing RGB images but also recover RAW images that are nearly the same as the original RAW data. The recovered RAW data are valuable for photographers and benefit a number of computer vision tasks such as HDR reconstruction~\cite{paris2011local}, image retouching~\cite{hu2018exposure}, and RAW compression. Here, the RAW images are visualized with bilinear demosaicing.}
\label{fig:teaser}
\vspace{-0.7em}
\end{figure*}

\section{Introduction}

Professional photographers can choose to process RAW images by themselves instead of RGB images to produce images with better visual effects as the RAW data captures unprocessed scene irradiance at each in 12-14 bits by a camera. Due to its linear relationship with scene irradiance, raw sensor data is also a better choice than RGB images for many image editing and computer vision tasks, such as photometric stereo, intrinsic image decomposition, image denoising, reflection removal, and image super resolution~\cite{bell2014intrinsic, brooks2019unprocessing, gharbi2016deep, Lei_2021_RFC, lei2020polarized, mildenhall2018burst, schwartz2018deepisp, shi2016benchmark,  zhang2019zoom}. 
However, accessing RAW images can be quite hard due to their memory-demanding property: RAW images may be discarded during the process of data storing, transferring, and sharing. In this paper, we are interested in the question: can users get access to the real RAW data without explicitly storing it? 

Due to the great advantages of RAW images, there have been many approaches to provide the mapping from sRGB images to their RAW counterparts~\cite{afifi2020cie, brooks2019unprocessing,  DBLP:conf/cvpr/LiuLCK0CH20, nguyen2018raw, 8658108, zamir2020cycleisp}. Nguyen et al.~\cite{nguyen2018raw} suggest explicitly storing the parameters of sRGB-RAW mapping functions into the JPEG metadata for the prospective RAW reconstruction. Brooks et al.~\cite{brooks2019unprocessing} use the prior information of the cameras (e.g., color correction matrices and digital gains) to reverse the ISP step-by-step. Another line of work~\cite{DBLP:conf/cvpr/LiuLCK0CH20, 8658108, zamir2020cycleisp} follows the inverse order of ISP and proposes learning-based methods to synthesize RAW data from sRGB images. However, these methods still rely on the underlying lossy in-camera ISP pipeline, and the recovered RAW images are inaccurate and may be different from the original ones. 

In this work, we propose a novel and effective learned solution by redesigning the camera image signal processing pipeline as an invertible one, which can be aptly called \textit{Invertible ISP (InvISP)}. Our learning-based InvISP enables rendering visually appealing RGB images in the forward process, and recovering nearly perfect quality raw sensor data from compressed RGB images through the inverse process. Our reconstructed RAW data is nearly identical with real RAW data and enables computer vision applications, such as image retouching and HDR reconstruction, as shown in Figure~\ref{fig:teaser}. 

Designing an invertible ISP is not a trivial task for at least three reasons. First, some steps in the traditional ISP, such as denoising, tone mapping, and quantization, can lead to inevitable information lost from wide-range (12-bit or 14-bit) raw sensor data to 8-bit RGB images. 
Second, the invertible ISP should not produce visual artifacts such as halo and ghosting artifacts~\cite{guarnieri2010high}. To render visually appealing sRGB images, denoising, demosaicing, color correction, white balance gain, tone mapping, and color enhancement must be designed carefully in ISP. Third, modern digital cameras store RGB images in the JPEG format, where the lossy compression process makes reconstructing high-quality RAW data highly challenging.

To overcome these challenges,  we take advantage of the inherent reversibility of normalizing-flow-based models~\cite{DBLP:conf/iclr/DinhSB17, DBLP:conf/nips/KingmaD18} and design both the RAW-to-RGB and RGB-to-RAW mapping in our invertible ISP with one single invertible neural network. We deeply analyze the properties of traditional ISP and design specific modules that can not only well approximate the camera ISP but also reconstruct almost identical RAW data with the camera RAW data. Specifically, we design our model with the composition of a stack of affine coupling layers and utilize the invertible 1 $\times$ 1 convolution as the learnable permutation function between the coupling layers. 
Besides, to empower our model to recover realistic RAW data from JPEG images, we integrate a differentiable JPEG simulator into our invertible neural network. We leverage the idea from Fourier transformation to replace the non-differentiable quantization step in JPEG compression. Thus, our end-to-end InvISP framework bypasses traditional ISP modules and minimizes the information loss for the RAW data and RGB image conversion. We bidirectionally train our network to optimize the RGB and RAW reconstruction process jointly. We experimentally prove that our framework can recover much better RAW data than state-of-the-art baselines without sacrificing the RGB reconstruction performance. 

To the best of our knowledge, our framework is the first attempt for RAW data reconstruction from the perspective of redesigning the camera ISP as an invertible one. Our method can address the information loss issue in ISP modules and is robust to the JPEG compression step. 
We demonstrate the effectiveness of our method on two DSLR cameras and show that our method outperforms state-of-the-art baselines to a large extent. Moreover, we also exhibit potential applications through RAW data compression, image retouching, and HDR reconstruction. 


\section{Related Work}
\mypara{RAW Image Reconstruction.} Recovering RAW from sRGB images has been well-studied~\cite{8658108, nguyen2018raw, afifi2020cie, brooks2019unprocessing, zamir2020cycleisp, mildenhall2018burst, DBLP:conf/cvpr/LiuLCK0CH20}. Nguyen et al.~\cite{nguyen2018raw} encode the parameters in ISP into JPEG metadata with 64KB overhead and use them to reconstruct RAW from JPEG images.  Brooks et al.~\cite{brooks2019unprocessing} propose to inverse the ISP pipeline step by step with camera priors. 
CIE-XYZ Net~\cite{afifi2020cie} proposes to recover RAW from sRGB images to the camera independent CIE-XYZ space. CycleISP~\cite{zamir2020cycleisp}
proposes to model the RGB-RAW-RGB data conversion cycle for synthesizing RAW from sRGB images. 
Unlike previous methods, we aim to fundamentally solve the RAW reconstruction problem by re-designing the camera ISP into an invertible one.

\mypara{Image Signal Processing (ISP).}
Image signal processing pipeline (ISP) aims at converting raw sensor data to human-readable RGB images~\cite{heide2014flexisp,Chen_2018_CVPR,Chen2019,schwartz2018deepisp,zhang2019zoom,Xu_2019_CVPR,DBLP:journals/tip/LiangCCZ21,klatzer2016learning,gharbi2016deep}. 
Heide et al.~\cite{heide2014flexisp} merge the steps in the traditional ISP pipeline to avoid the accumulative error.
Gharbi et al.~\cite{gharbi2016deep} propose a method with end-to-end networks to learn RAW demosaicing and denoising jointly. Hasinoff et al.~\cite{hasinoff2016burst} propose a low-light imaging system for mobile devices.
Other works~\cite{Chen_2018_CVPR, schwartz2018deepisp} focus on learning low-light enhancement ISP pipelines with CNNs. 
Zhang et al. ~\cite{zhang2019zoom} process RAW for super-resolution task with U-net~\cite{ronneberger2015u} to preserve high-frequency information. CameraNET ~\cite{DBLP:journals/tip/LiangCCZ21} splits the ISP into two learning stages for CNN. Unlike the encoder-decoder style network adopted in previous work, we demonstrate that invertible neural networks own great potential for ISP pipeline and enable accurate RAW reconstruction.

\mypara{Invertible Neural Networks. }
Normalizing flow-based invertible neural networks~\cite{DBLP:conf/icml/HoCSDA19,DBLP:conf/nips/KingmaD18,DBLP:journals/corr/DinhKB14,DBLP:conf/iclr/DinhSB17} have become a popular choice in image generation tasks. Normalizing flow transforms a simple posterior distribution to a complex real-world distribution through a series of invertible transformations. NICE \cite{DBLP:journals/corr/DinhKB14} is the first learning-based normalizing flow framework with the proposed additive coupling layers. RealNVP \cite{DBLP:conf/iclr/DinhSB17} modifies the additive coupling layer to both multiplication and addition, and composes the coupling layer in an alternating pattern such that all the inputs can be altered with equal chance. Kingma et al.~\cite{DBLP:conf/nips/KingmaD18} propose ActNorm layer and generalize channel-shuffle operations with invertible $1\times1$ convolution. Flow++ \cite{DBLP:conf/icml/HoCSDA19} modifies the affine coupling layer to logistics mixture CDF coupling flows and applies self-attention module.

\section{Traditional ISP analysis}
Modern digital cameras apply a series of operations, which form the image signal processing pipeline (ISP), to render RAW data to human-readable RGB images. These operations include white balance, demosaicing, denoising, color space transformation, tone mapping, and others~\cite{karaimer2016software}. 
Traditionally, every step of an ISP needs labor-intensive tuning for specific cameras, and inverting the traditional ISP steps is quite challenging.
In this section, we analyze the existing modules with information loss in the traditional ISP. We show that the lossy steps in traditional ISP restrict the RAW reconstruction performance of a series of works~\cite{nguyen2018raw, brooks2019unprocessing, zamir2020cycleisp} that aim at synthesizing RAW from sRGB images. 
Different from previous works, we re-design the ISP into an end-to-end invertible one that can bypass the traditional modules to minimize information loss during the RAW data and JPEG image conversion, which further enables recovering high-quality RAW data.

\begin{figure}[t]
\centering
\begin{tabular}{c@{}}
\includegraphics[width=0.98\linewidth]{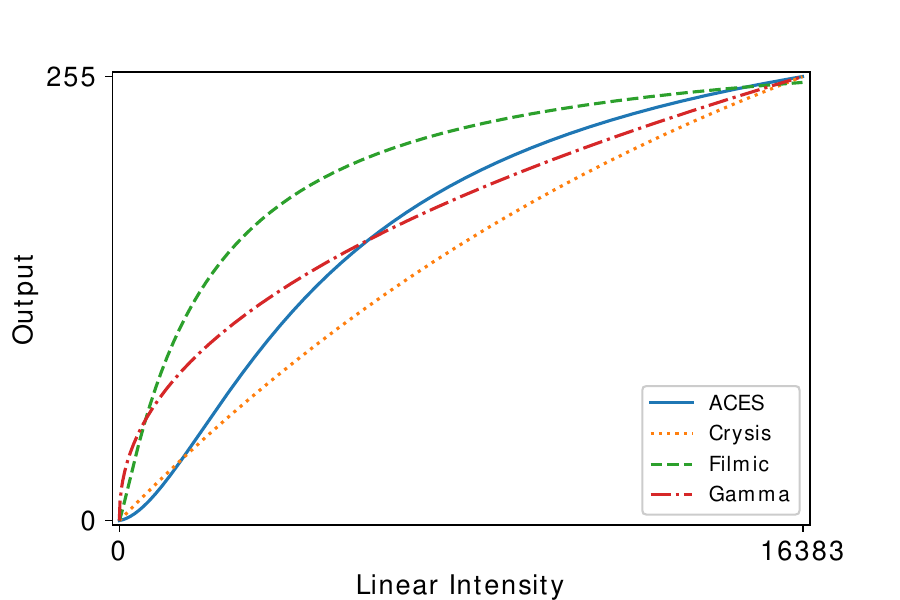}\\
\end{tabular}
\caption{Some popular tone mapping curves used in games and industries ~\cite{reinhard2002photographic, 6811423}. Although the tone mapping function itself is lossless, the following quantization causes a great loss of information in over-exposed and under-exposed pixels. For instance, in a 14-bit linear RAW image, the pixel intensity lies in [16313, 16383] will all be quantized to the maximum pixel intensity 255 of an 8-bit RGB image. }
\label{fig:tone}
\vspace{-0.7em}
\end{figure}

\mypara{Quantization and tone mapping. }
Some ISP steps like demosaicing and gamma compression may involve float-point operations, and thus quantization is inevitable to transform the data into the integer type. 
For instance, the rounding function can bring $(-0.5, 0.5)$ intensity error to a pixel in theory. 
In the context of ISP, however, the tone mapping step can enlarge the intensity error much greater than $\pm$0.5. 
The tone mapping curve is usually designed as S-curve that compresses the high-intensity value and low-intensity value more than mid-intensity values~\cite{reinhard2002photographic, 6811423}. 
As illustrated in Figure~\ref{fig:tone}, for a 14-bit raw image, gamma compression makes pixel intensity at $[16313, 16383]$ all be rounded to the max intensity 255 after normalized to $(0,255)$. This step may cause a 0.004 RMSE error at this single pixel. Thus, it is challenging for existing works~\cite{nguyen2018raw, brooks2019unprocessing, zamir2020cycleisp} to directly synthesize the 14-bit RAW data from its 8-bit sRGB counterparts, especially at the over-exposed regions. We show the comparison of our recovered RAW with previous works in Figure~\ref{fig:cycleisp}. Our method can preserve much more detail of RAW data, even at high-intensity pixels. 

\begin{figure*}[t!]
\centering
\includegraphics[width=1.0\textwidth]{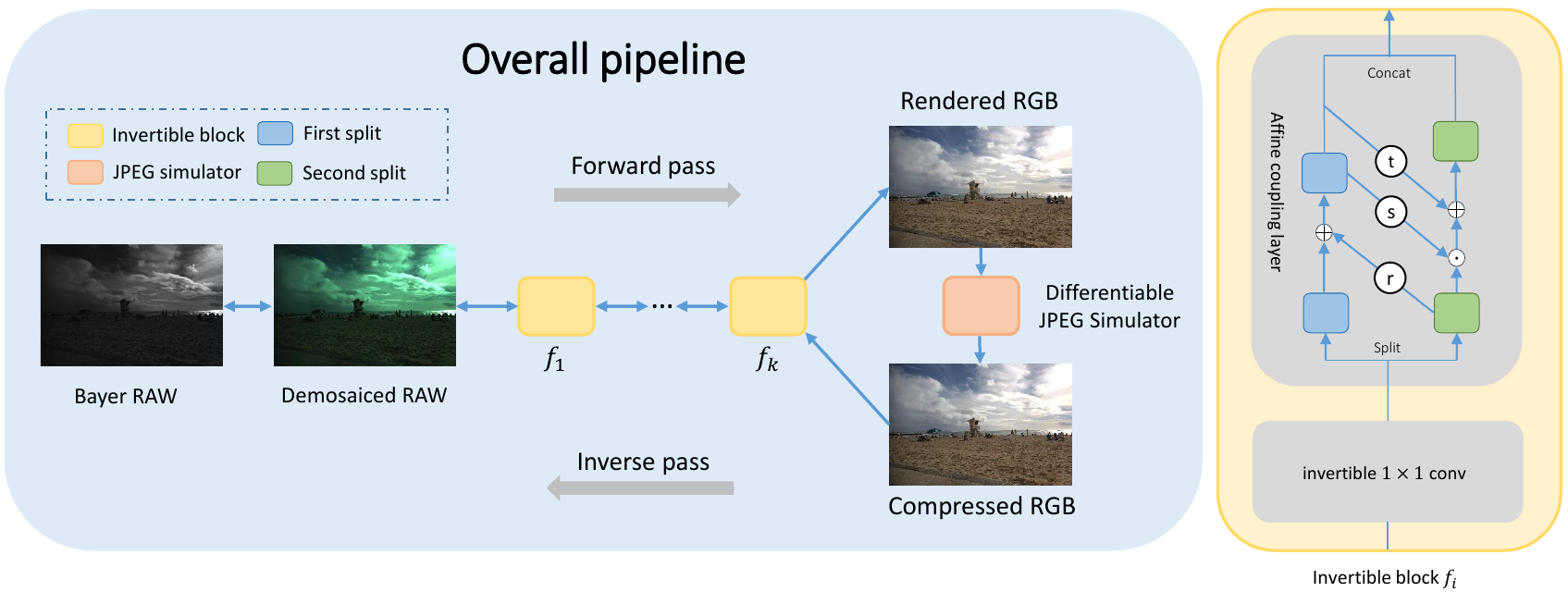}
\vspace{-0.7em}
\caption{Our Invertible ISP (InvISP) framework. InvISP is composed of both forward and inverse passes. In the forward pass, the Bayer RAW is first bilinearly demosaiced and then transformed to an RGB image by a stack of bijective functions $\{f_i\}_{i=0}^k$. Our model integrates a differentiable JPEG simulator to account for compression information lost. During the training time, to invert the ISP, the backward pass takes a compressed RGB image as input and reverses all the bijective functions and the bilinear demosaicing to obtain the original RAW image. Note that the backward pass takes real JPEG images as input at test time. We illustrate the details of the invertible block on the right. $r$, $s$, and $t$ are transformations defined in the bijective functions $\{f_i\}_{i=0}^k$. }
\label{fig:overview}
\vspace{-0.7em}
\end{figure*}

\mypara{Out-of-range value clipping.}
Value clipping is a common step to normalize the raw value within a reasonable range, which may happen after color space transformation, demosaicing, denoising, and tone mapping~\cite{DBLP:conf/cvpr/AbdelhamedLB18, foi2009clipped, plotz2017benchmarking, endo2017deepReverseTone}. Most commonly used value clipping operation is like $\text{min}(\text{max}(x, 0), 1)$, which will discard the out-of-range pixels at over- and under-exposed regions. Note that this restricts the image capacity for further adjustment. 
Moreover, traditional ISPs are manually tuned in isolation by experts, which accumulates the clip error among ISP steps to bring further information lost. Our end-to-end pipeline jointly optimizes all the ISP steps and alleviates the clip error accumulation problem to recover more realistic RAW images. 

\mypara{JPEG compression.}
Modern digital cameras store RGB images in JPEG format, whose information loss further brings challenges to RAW image reconstruction. 
JPEG encoding pipeline consists of four main steps:  color space transformation, discrete cosine transformation (DCT), quantization, and entropy encoding~\cite{pennebaker1992jpeg}.
In reality, quantization is the only lossy and non-differentiable step in JPEG compression. 
Note that the JPEG information loss is quite hard to reverse. 
Thus we take a compromised step by integrating the JPEG compression procedure into our network optimization process to alleviate the information loss. To achieve this, we design a differentiable JPEG simulator by carefully simulating the JPEG compression procedure and replacing the quantization step with differentiable Fourier transformations. 

\section{Method} 
\subsection{Invertible Image Signal Processing (InvISP)} 
We denote the RAW data space as $\mathcal{X}$ and sRGB data space as $\mathcal{Y}$. Our goal is to find the invertible and bijective function which can map the data point from RAW data space to sRGB data space, denoted as $f: \mathcal{X} \rightarrow \mathcal{Y}$. 
To achieve this, classical neural networks need two separate networks to approximate $\mathcal{X} \rightarrow \mathcal{Y}$ and $\mathcal{Y} \rightarrow \mathcal{X}$ mappings respectively, which leads to inaccurate bijective mapping and may accumulate the error of one mapping into the other.
We take an alternative method and use the affine coupling layers in~\cite{DBLP:conf/iclr/DinhSB17, DBLP:conf/nips/KingmaD18} to enable invertibility of one single network. We design our invertible ISP with the compostition of a stack of invertible and tractable bijective functions $\{f_i\}_{i=0}^k$, i.e. $f = f_0 \circ f_1 \circ f_2 \circ \cdots \circ f_k$. 
For a given observed data sample $\mathbf{x}$, we can derive the transformation to target data sample $\mathbf{y}$ through 
\begin{align}
    \mathbf{y} &= f_0 \circ f_1 \circ f_2 \circ \cdots \circ f_k(\mathbf{x}), \\ 
    \mathbf{x} &= f_k^{-1} \circ f_{k-1}^{-1} \circ \cdots \circ f_0^{-1}(\mathbf{y}). 
\end{align}

The bijective model $f_i$ is implemented through affine coupling layers. In each affine coupling layer, given a $D$ dimensional input $\mathbf{m}$ and $d < D$, the output $\mathbf{n}$ is calculated as 
\begin{align}
    \mathbf{n}_{1: d} &=\mathbf{m}_{1: d}, \label{coup_1} \\ 
    \mathbf{n}_{d+1: D} &=\mathbf{m}_{d+1: D} \odot \exp \left(s\left(\mathbf{m}_{1: d}\right)\right)+t\left(\mathbf{m}_{1: d}\right),
\end{align}
where $s$ and $t$ represent scale and translation functions from $R^d \mapsto R^{D-d}$, and $\odot$ is the Hadamard product. 
Note that the scale and translation functions are not necessarily invertible, and thus we realize them by neural networks. 

As stated in~\cite{DBLP:conf/iclr/DinhSB17}, the coupling layer leaves some input channels unchanged, which greatly restricts the representation learning power of this architecture. To alleviate this problem, we firstly enhance~\cite{DBLP:conf/eccv/XiaoZLWHKBLL20} the coupling layer (\ref{coup_1}) by 
\begin{align}
    \mathbf{n}_{1: d} &=\mathbf{m}_{1: d} + r(\mathbf{m}_{d+1: D}),
\end{align}
where $r$ can be arbitrary function from $R^{D-d} \mapsto R^d$.  The inverse step is easily obtained by 
\begin{align}
    \mathbf{m}_{d+1: D}&=\left(\mathbf{n}_{d+1: D}-t\left(\mathbf{n}_{1: d}\right)\right) \odot \exp \left(-s\left(\mathbf{n}_{1: d}\right)\right), \\
    \mathbf{m}_{1: d} \quad&=\mathbf{n}_{1: d} - r(\mathbf{m}_{d+1: D}).
\end{align}

Next, we utilize the invertible 1 $\times$ 1 convolution proposed in~\cite{DBLP:conf/nips/KingmaD18} as the learnable permutation function to reverse the order of channels for the next affine coupling layer.

We remove the spatial checkerboard mask as it brings no evident performance improvement~\cite{DBLP:conf/nips/KingmaD18}. We follow the implementation of~\cite{Chen_2018_CVPR} and disable batch normalization~\cite{DBLP:conf/icml/IoffeS15} and weight normalization used in~\cite{DBLP:conf/iclr/DinhSB17}. 
For our image-to-image translation task, we directly learn the RAW-to-RGB mapping without explicitly modeling the latent distribution to stabilize the training process.

Note that the input size of invertible neural networks must be identical to the output size. Thus, we take the bilinear demosaiced RAW data as input, which will not destroy the RAW data quality, and reversing the bilinear demosaicing is trivial~\cite{brooks2019unprocessing}. For the affine coupling layer, we split the input into two parts. We note that although three-channel input cannot be split evenly, the invertible 1 $\times$ 1 convolution ensures that unchanged components are updated in the next invertible block. Thus R, G, and B channels are still treated equally. We also do an online gamma correction (\textit{i.e.} without storing on disk) to RAW data to compress the dynamic range for faster convergence speed. 

The forward pass of our InvISP produces the sRGB images, and the reverse pass aims at recovering realistic RAW data. We conduct bi-directional training with $L_1$ loss to optimize our framework. 

\begin{align}
    L = ||f(\mathbf{x}) - \mathbf{y}||_{1} + \lambda ||f^{-1}(\mathbf{y}) - \mathbf{x}||_{1}, 
\end{align}
where $\lambda$ is the hyper-parameter used to balance the accuracy between RGB and RAW reconstruction. We set $\lambda$ to 1 in our main experiments.

\begin{figure}
\centering
\begin{tabular}{c@{\hspace{0.5mm}}c@{\hspace{0.5mm}}c@{}}
\includegraphics[width=0.8\linewidth]{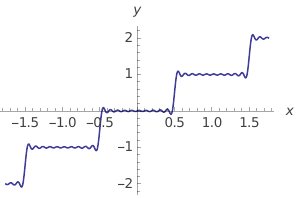}\\
\end{tabular}
\vspace{1mm}
\caption{The curve of our approximation rounding function for quantization in our differentiable JPEG simulator.}
\label{fig:distort}
\vspace{-1.5mm}
\end{figure}

\begin{table*}[ht]
\centering
\setlength{\tabcolsep}{3mm}
\ra{1.25}
\begin{tabular}{l@{\hspace{18mm}}c@{\hspace{6mm}}c@{\hspace{8mm}}c@{\hspace{18mm}}c@{\hspace{6mm}}c@{\hspace{8mm}}c}
\toprule
& \multicolumn{3}{@{\hspace{-19mm}}c}{NIKON D700} & \multicolumn{3}{@{\hspace{-2mm}}c}{Canon EOS 5D}  \\
Method &\multicolumn{2}{@{\hspace{-5mm}}c}{RGB}  & RAW  & \multicolumn{2}{@{\hspace{-5mm}}c}{RGB}  & RAW   \\
 & PSNR  & SSIM  & PSNR  & PSNR & SSIM  & PSNR  \\
\midrule
UPI~\cite{brooks2019unprocessing} & - & - & 30.12 & - & - & 26.31\\
CycleISP~\cite{zamir2020cycleisp} & - & - & 30.19 & - & - & 34.48\\
\midrule
InvGrayscale~\cite{xia2018invertible} & 24.13 & 0.8258 & 33.28 & 28.22 & 0.8714 & 38.00\\
U-net~\cite{Chen_2018_CVPR} & 36.48 & 0.9342 & 41.17 & 33.44 & 0.8893 & 41.14\\
\midrule
Ours (w/o JPEG simulation) & 37.44 & 0.9309 & 44.19 & 33.45 & 0.8923 & 45.73 \\
Ours (JPEG with DSQ~\cite{gong2019differentiable}) & 37.44 & 0.9467 & \textbf{45.25} & 33.15 & 0.8946 & 48.22\\
Ours (JPEG with Fourier) & \textbf{37.47} & \textbf{0.9473} & 45.23 & \textbf{33.61} & \textbf{0.9007} & \textbf{48.57} \\
\bottomrule
\end{tabular}
\vspace{1mm}
\caption{Quantitative evaluation among our model and baselines. Various perceptual metrics show that our proposed ISP model outperforms all the baselines. Our method with JPEG simulation using proposed Fourier quantization outperforms the other two alternative models. }
\label{table:quant_other}
\end{table*}

\begin{figure*}[t]
\centering
\begin{tabular}{c@{\hspace{0.9mm}}c@{\hspace{0.7mm}}c@{\hspace{0.7mm}}c@{\hspace{0.7mm}}c@{\hspace{0.7mm}}c@{\hspace{0.7mm}}c@{}}
\rotatebox{90}{\small \hspace{18mm}}&
\includegraphics[width=0.19\linewidth]{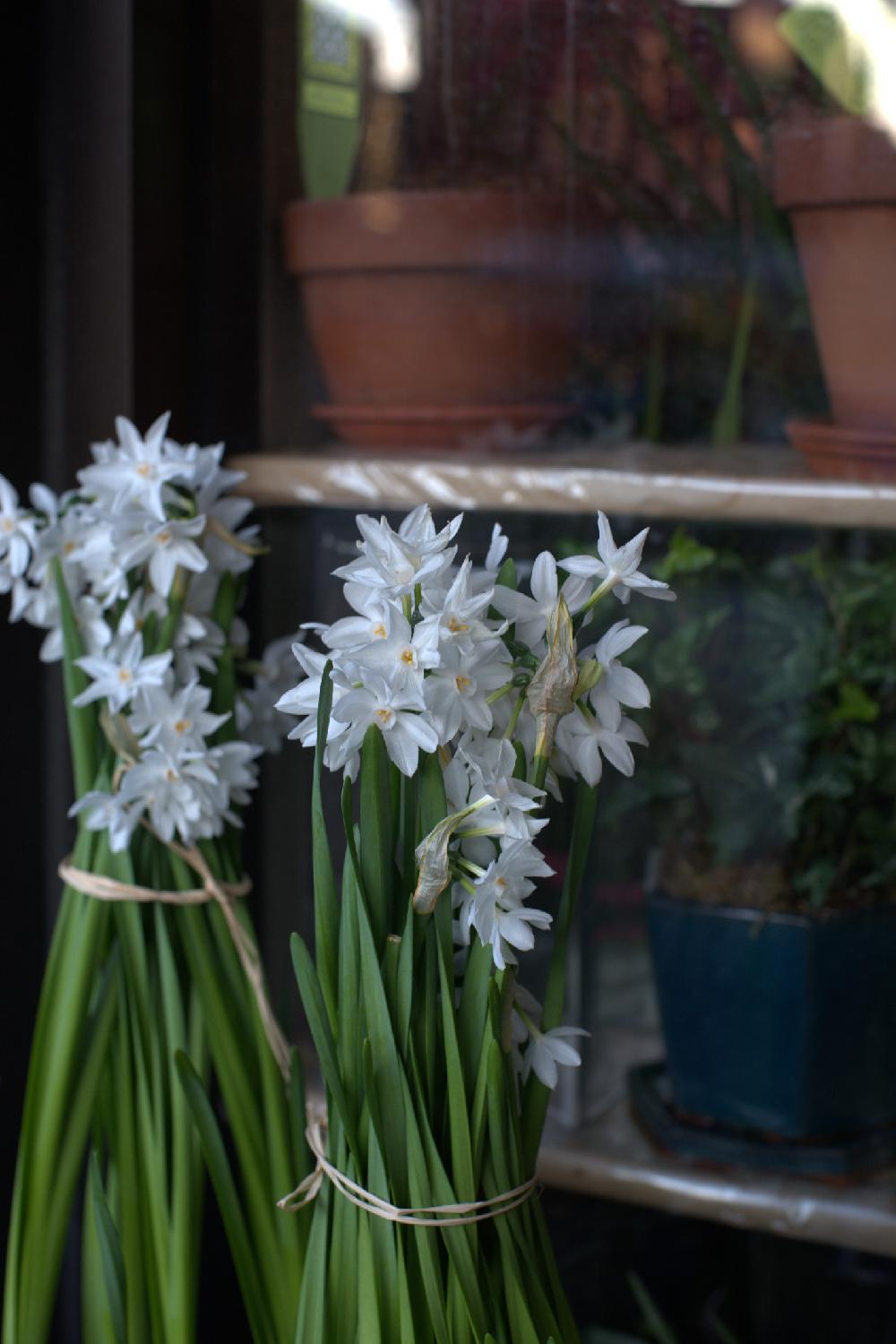}&
\includegraphics[width=0.19\linewidth]{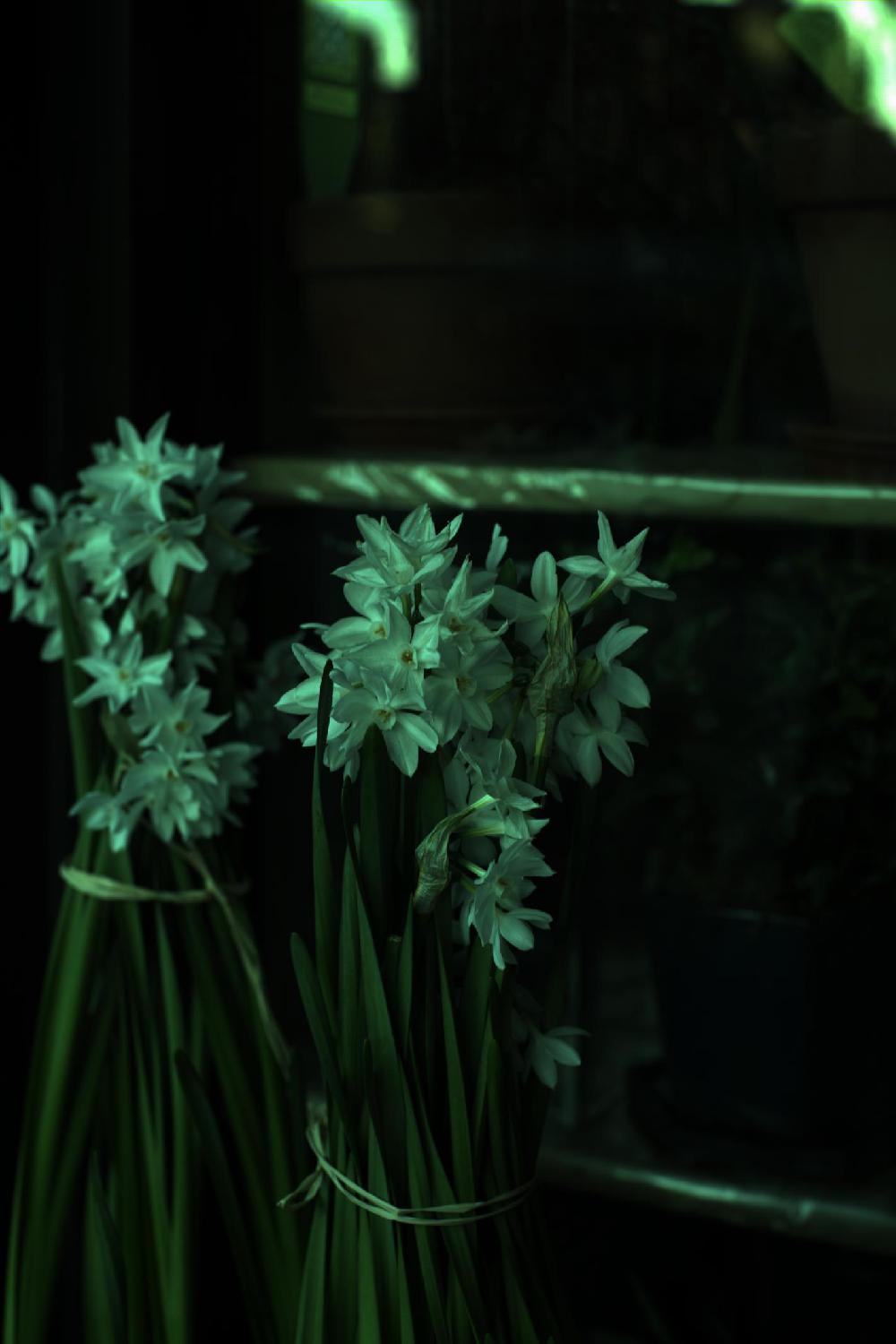}&
\includegraphics[width=0.19\linewidth]{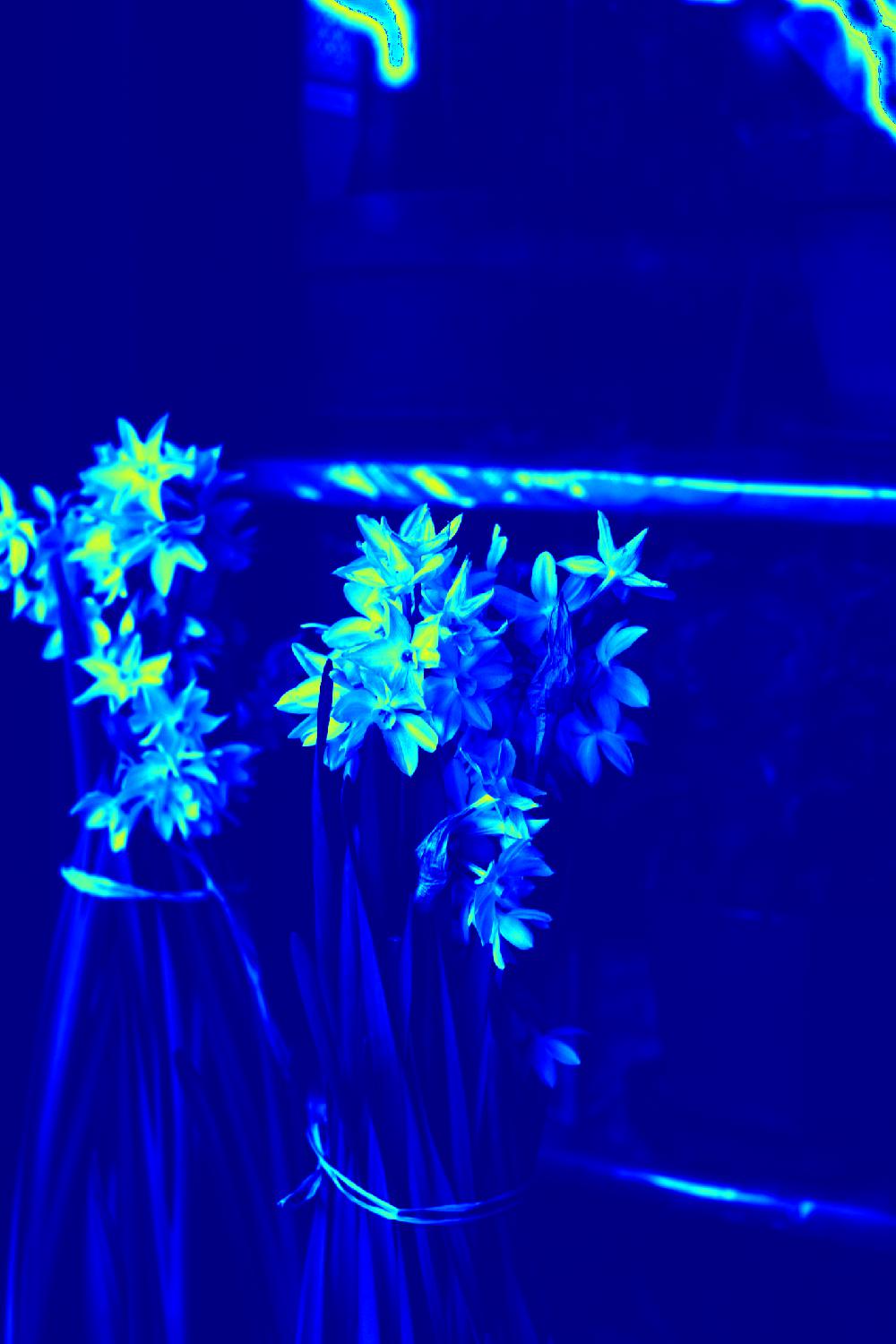}&
\includegraphics[width=0.19\linewidth]{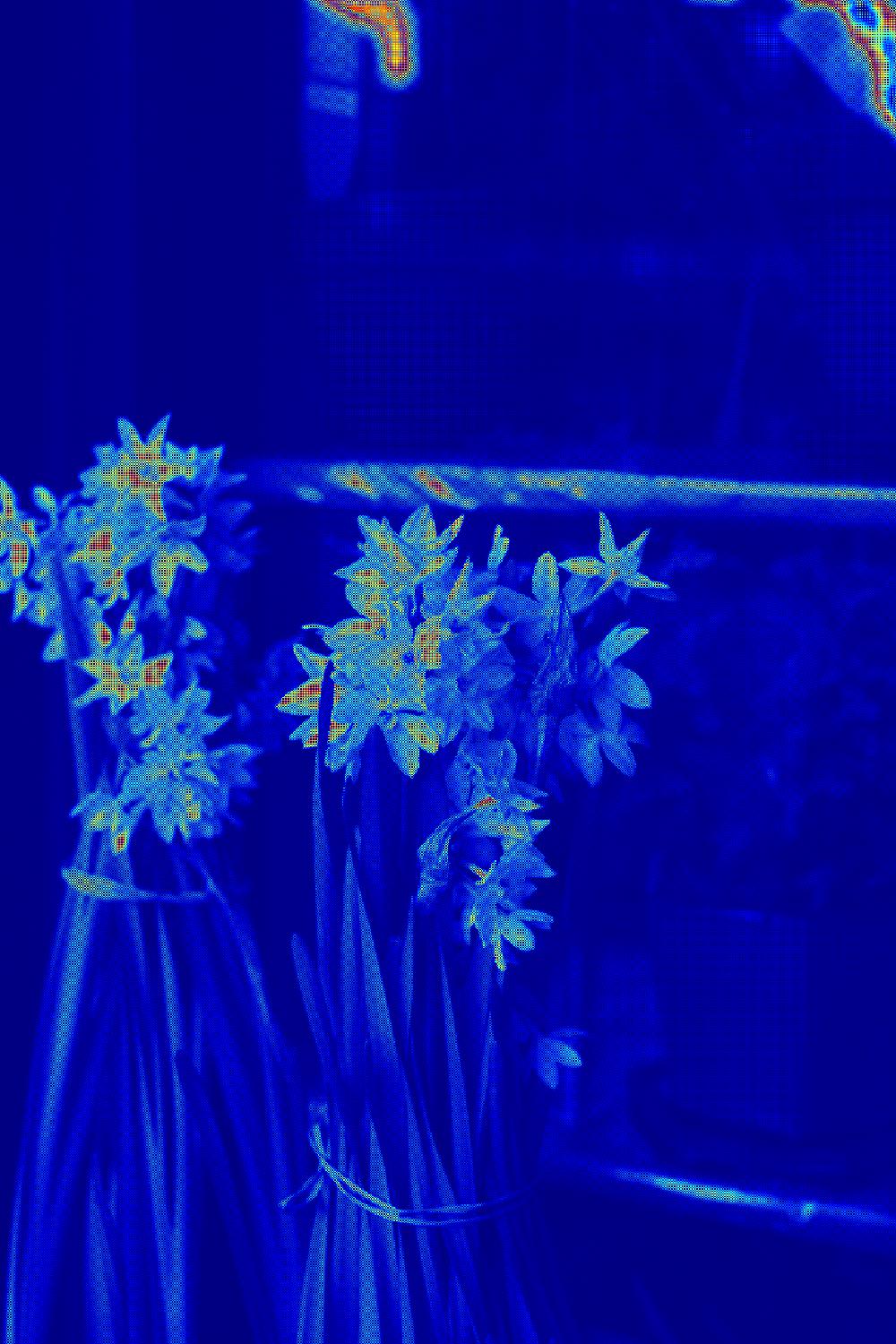}&
\includegraphics[width=0.19\linewidth]{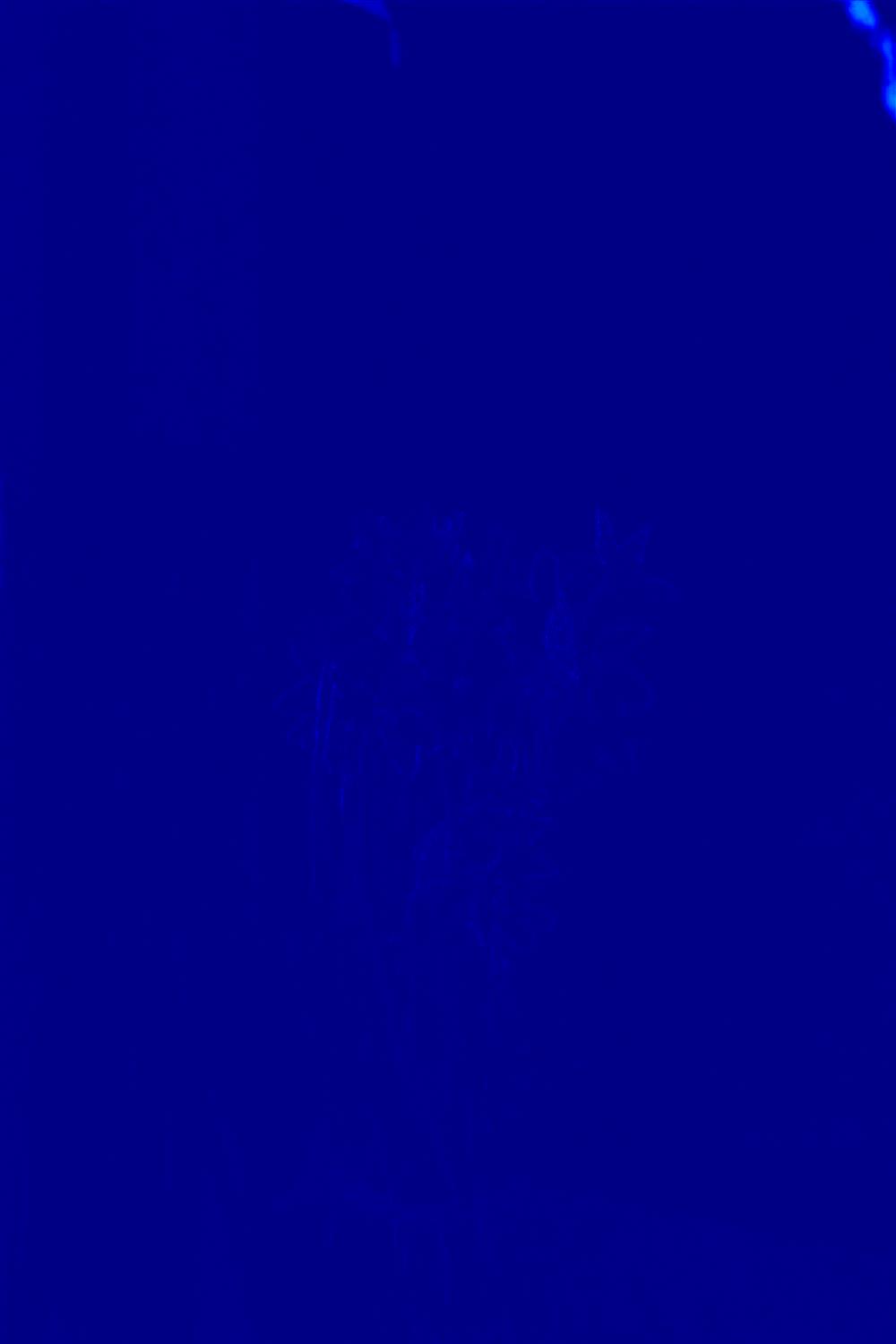}&\\
\rotatebox{90}{\small \hspace{25mm}}&
\includegraphics[width=0.19\linewidth]{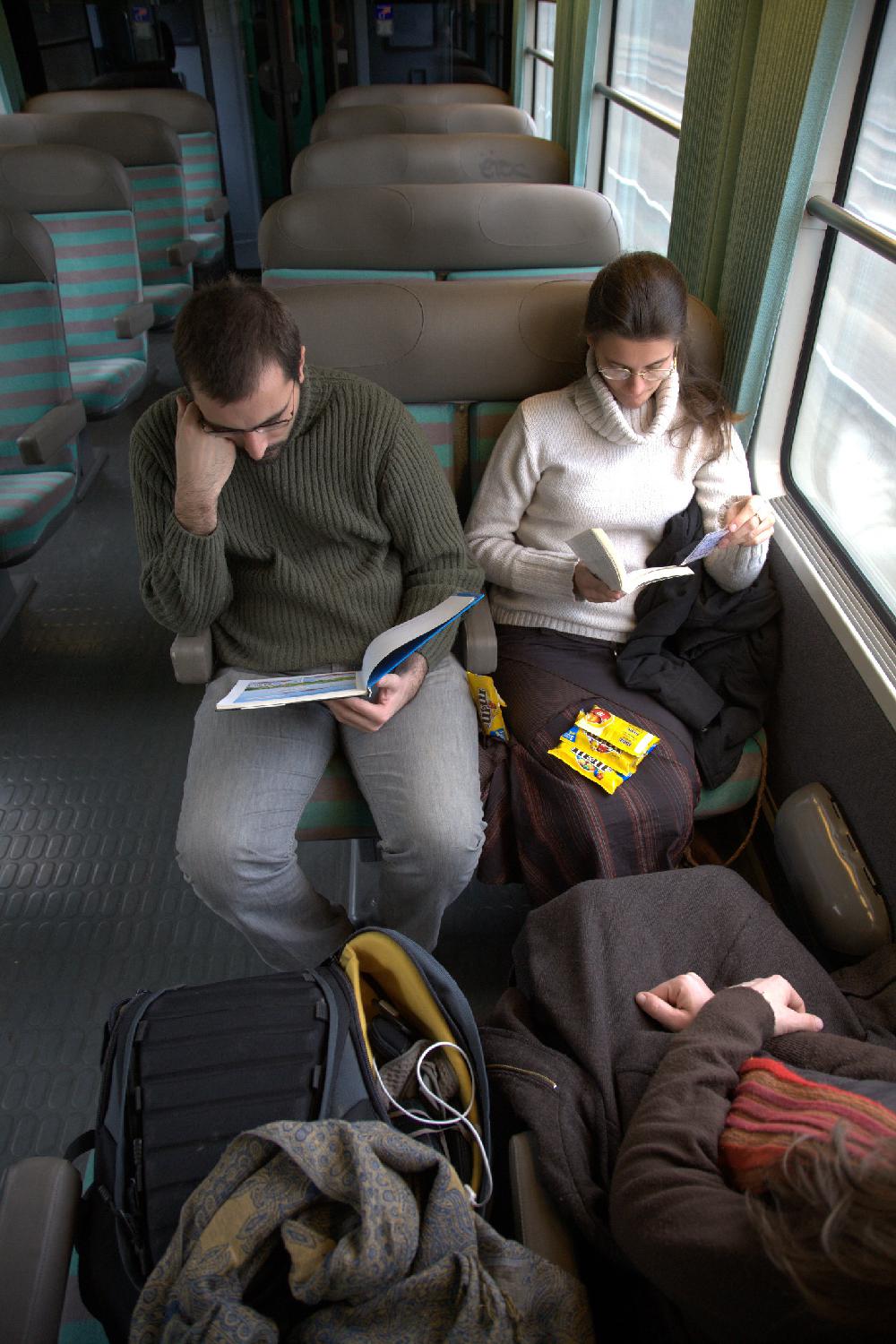}&
\includegraphics[width=0.19\linewidth]{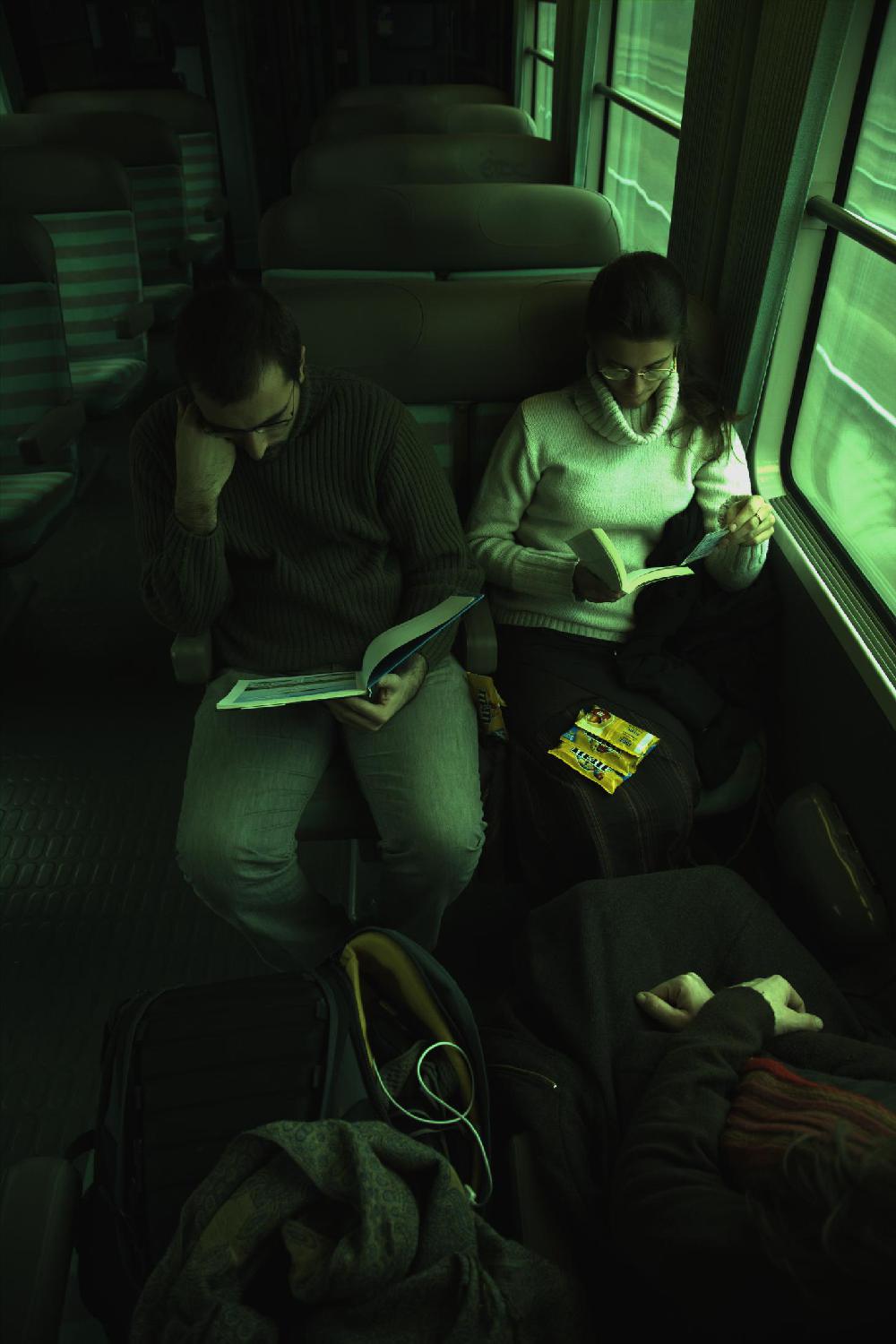}&
\includegraphics[width=0.19\linewidth]{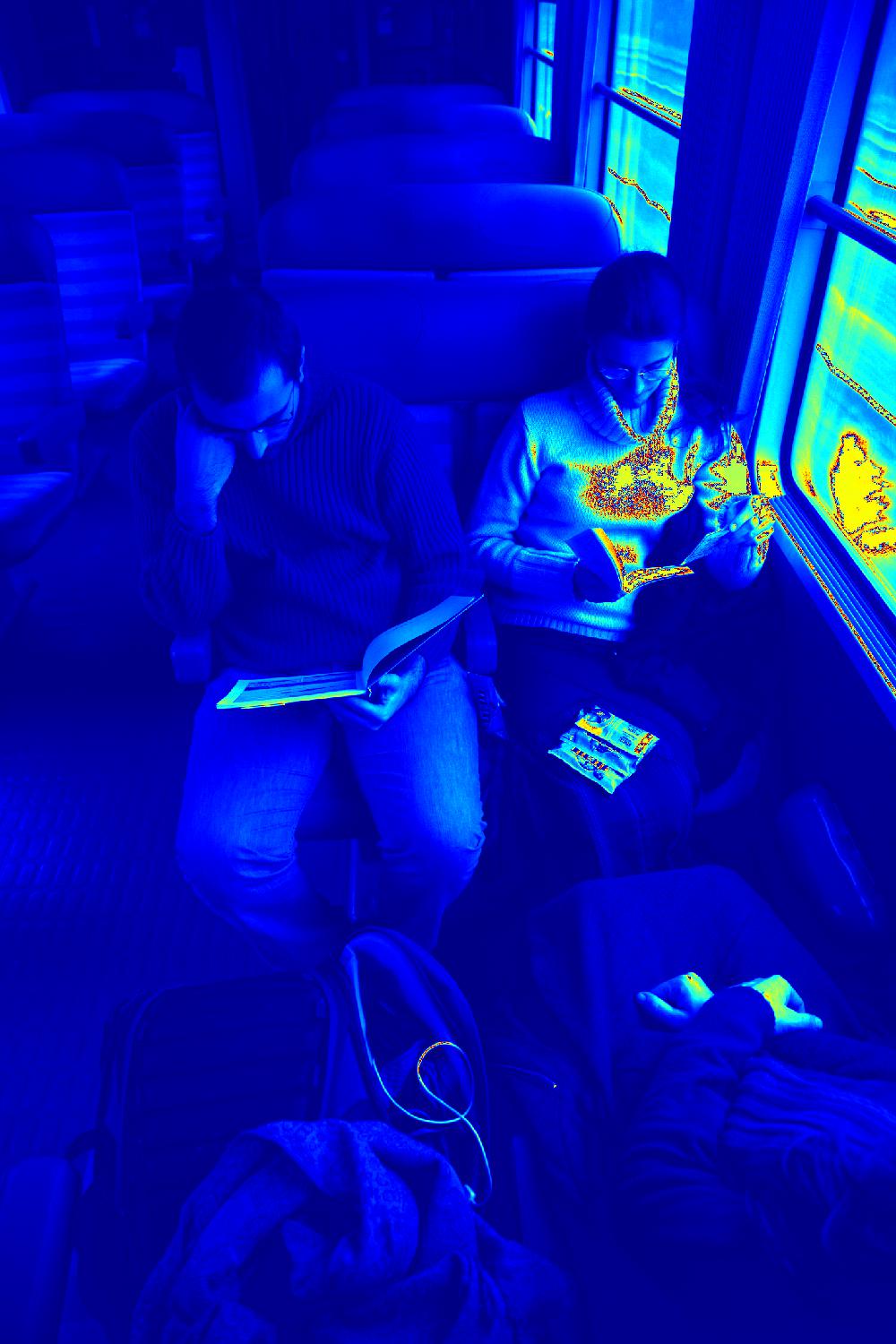}&
\includegraphics[width=0.19\linewidth]{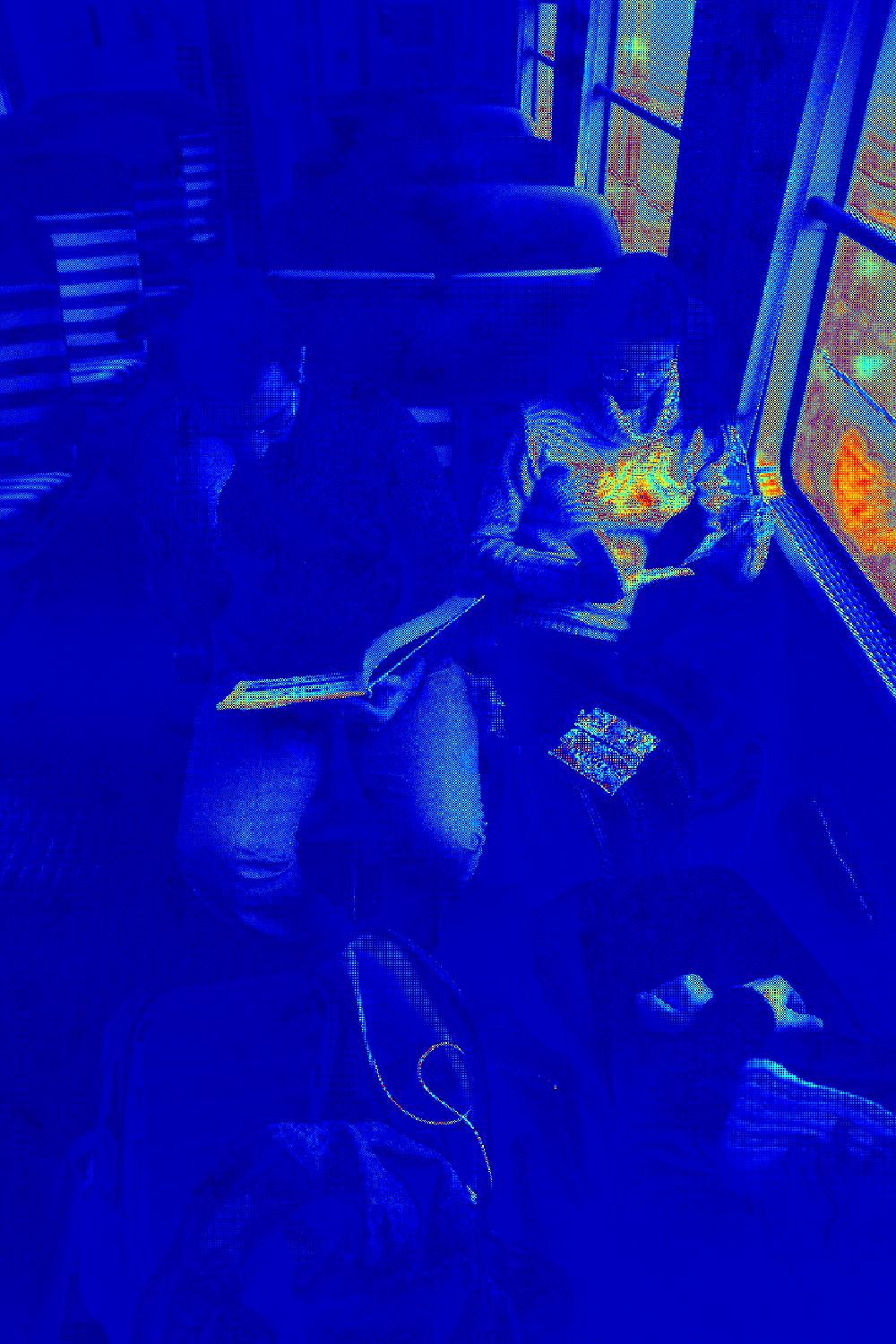}&
\includegraphics[width=0.19\linewidth]{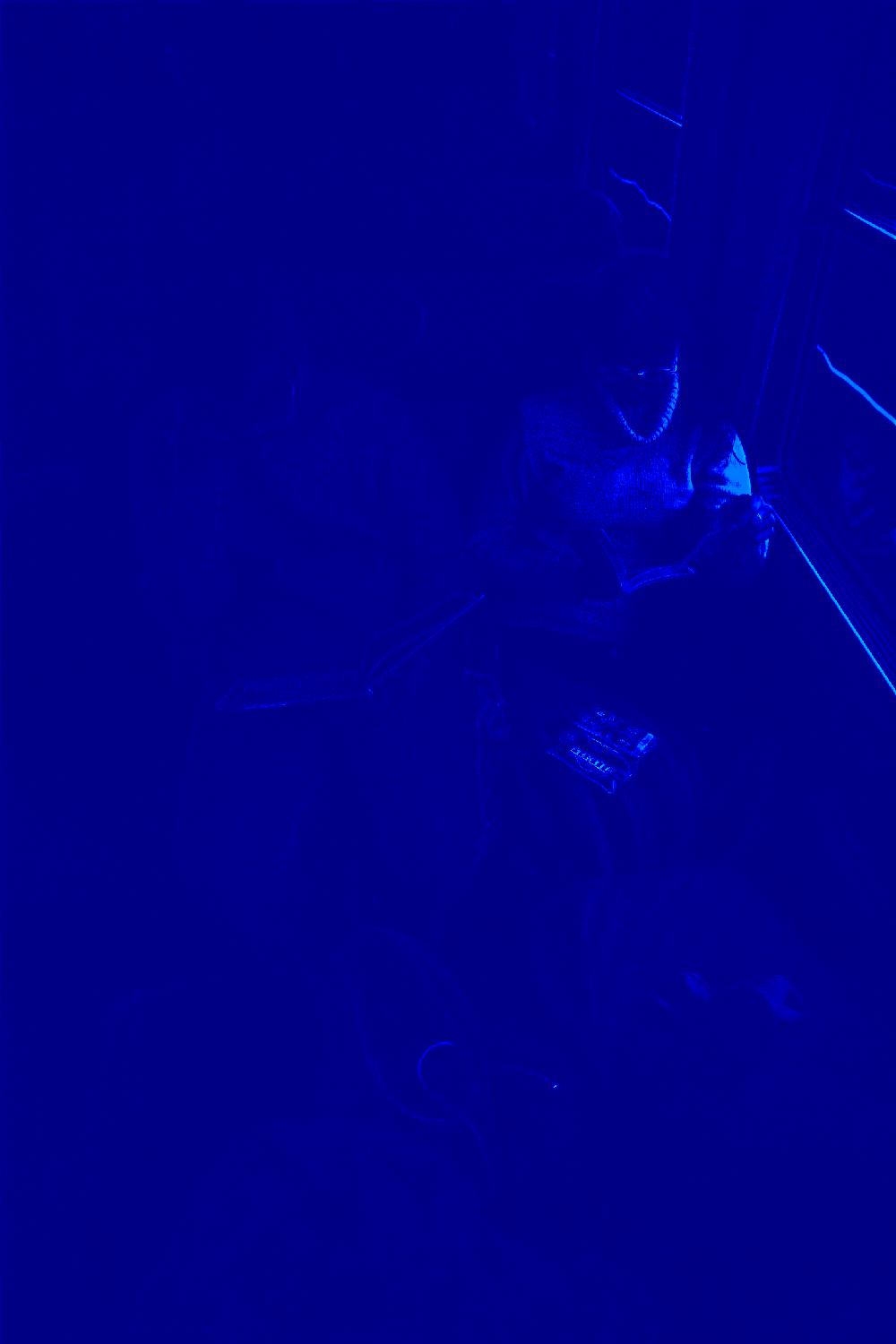}&\\
&Ground-truth RGB &Ground-truth RAW  &UPI RAW~\cite{brooks2019unprocessing} & CycleISP RAW~\cite{zamir2020cycleisp} & Our RAW \\
\end{tabular}
\vspace{1mm}
\caption{The qualitative comparison among UPI~\cite{brooks2019unprocessing}, CycleISP~\cite{zamir2020cycleisp} and our method. UPI and CyleISP synthesize RAW data from 8-bit compressed RGB, which is inevitable to suffer from the information loss of traditional ISP. Unlike theirs, our model forms a RAW-RGB-RAW cycle and is inherently reversible to recover the realistic RAW image. The GT RAW image is visualized through bilinear demosaicing, and other RAW images are visualized through error maps. This figure is best viewed in the electronic version.}
\label{fig:cycleisp}
\vspace{-0.7em}
\end{figure*}

\subsection{Differentiable JPEG Simulator}

Our goal is to train a robust invertible ISP that can tolerate the distortion by JPEG compression to recover accurate RAW.  
However, the JPEG compression algorithm is not differentiable, which can not be directly integrated into our end-to-end framework. Thus, we propose a differentiable JPEG simulator to enable our network robust to the JPEG compression through the optimization process. Since entropy encoding is lossless and goes after quantization, we skip this step and only simulate color space transformation, DCT, and quantization steps. 

To simulate the DCT process, we compute the DCT coefficients and split the input into $8 \times 8$ blocks. Then each block is multiplied by DCT coefficients to get the DCT map. In JPEG compression, the DCT map is divided by quantization tables and rounding to the integer type.
Since the rounding function is not differentiable, we design a differentiable rounding function base on the Fourier series, which can be defined as: 
\begin{align}
    Q(I)=I-\frac{1}{\pi}\sum_{k=1}^{K}\frac{(-1)^{k+1}}{k} \text{sin}(2 \pi k I),
\end{align}
where $I$ is the input map after divided by quantization tables in JPEG compression, and $K$ is used for the tradeoff between approximation accuracy and computation efficiency. As $K$ increases, the simulation function is closer to the real round function, but the running time will also increase. We empirically set $K$ to 10. The rounding process is illustrated in Figure~\ref{fig:distort}.

In the decoding phase of JPEG compression, $I$ is multiplied by the quantization table. The inverse DCT and color space transformation are then applied to reconstruct the simulated JPEG images.

\mypara{Discussion. } Differentiable rounding function is widely used in network quantization research. To fairly prove the effectiveness of our proposed rounding function, we also compare with the rounding function in~\cite{gong2019differentiable}, as shown in Table~\ref{table:quant_other}. Our method can achieve a better balance between RGB rendering and RAW reconstruction.

\begin{figure*}[!]
\centering
\hspace*{-2mm}
\begin{tabular}{c@{\hspace{0.7mm}}c@{\hspace{0.7mm}}c@{\hspace{0.7mm}}c@{\hspace{0.7mm}}c@{\hspace{0.7mm}}c@{}}
\rotatebox{90}{\small \hspace{6mm} RAW image}&
\includegraphics[width=0.24\linewidth]{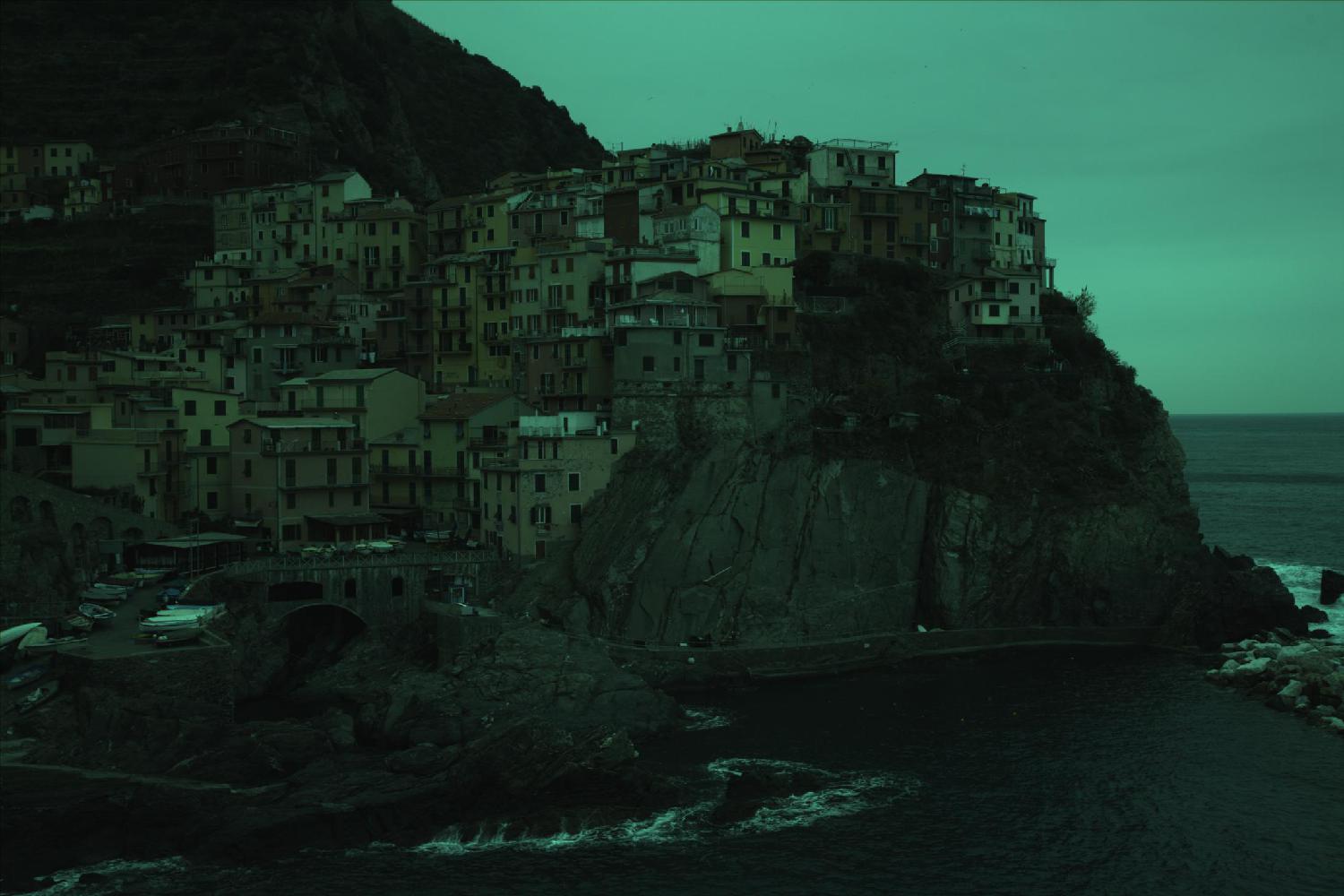}&
\includegraphics[width=0.24\linewidth]{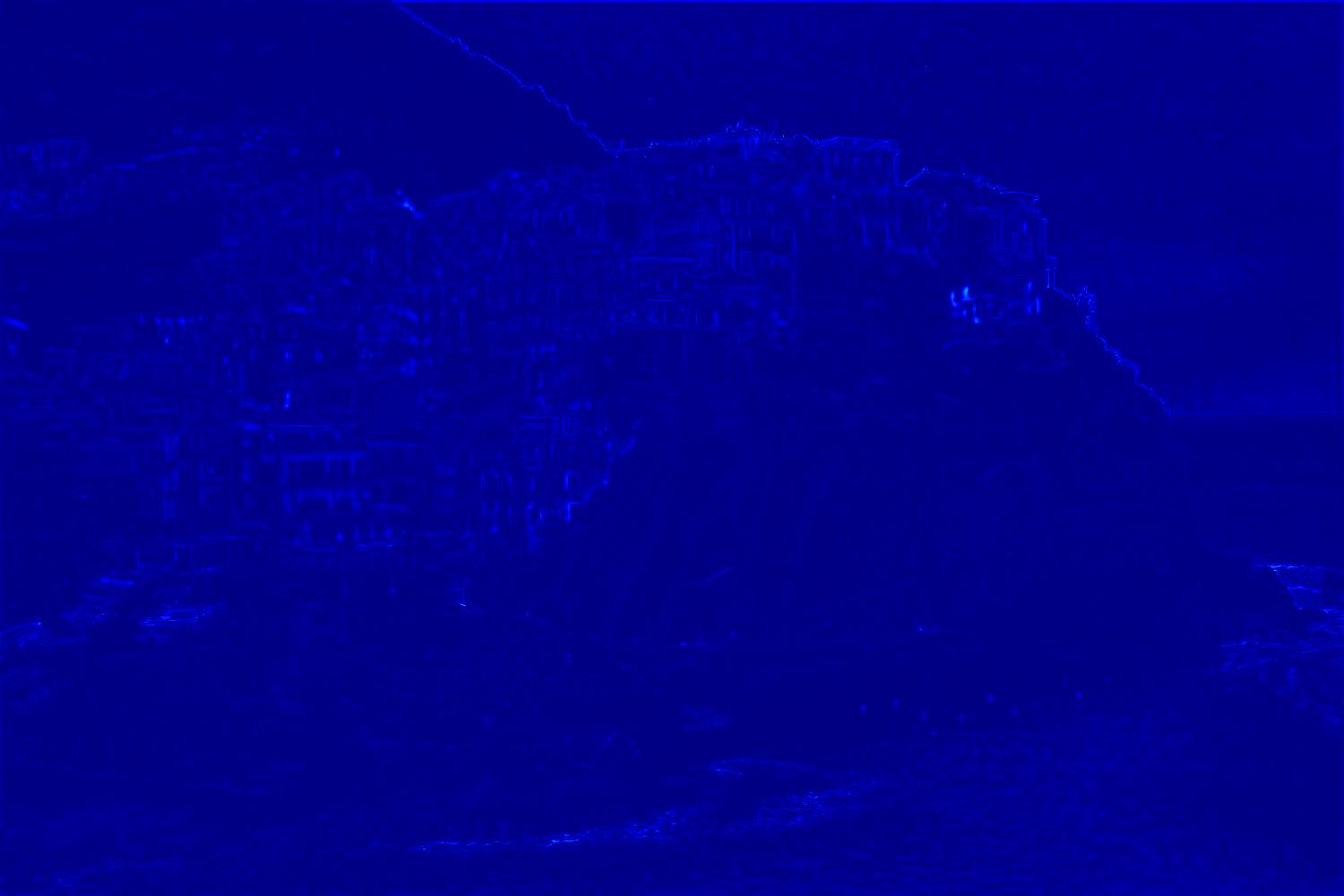}&
\includegraphics[width=0.24\linewidth]{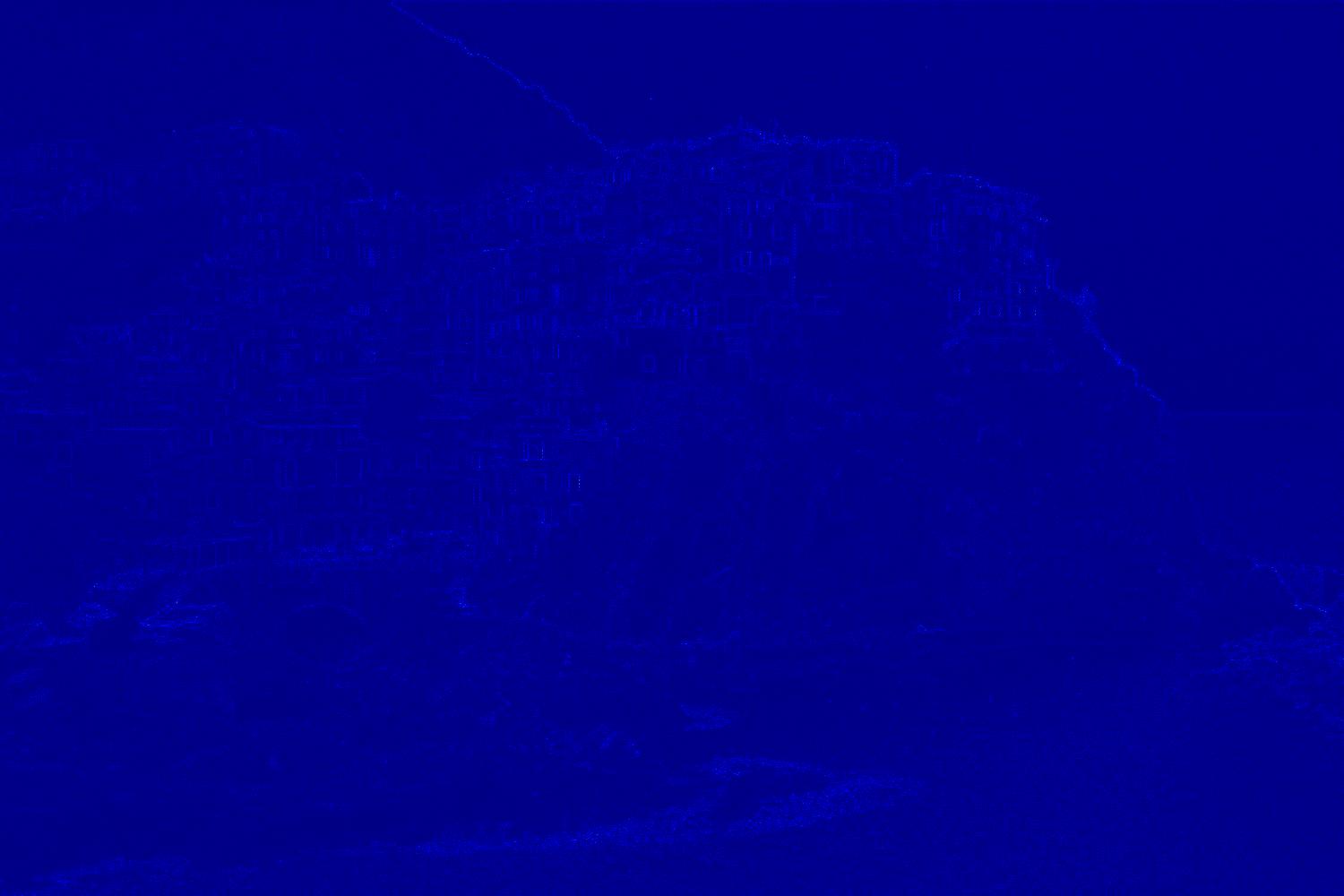}&
\includegraphics[width=0.24\linewidth]{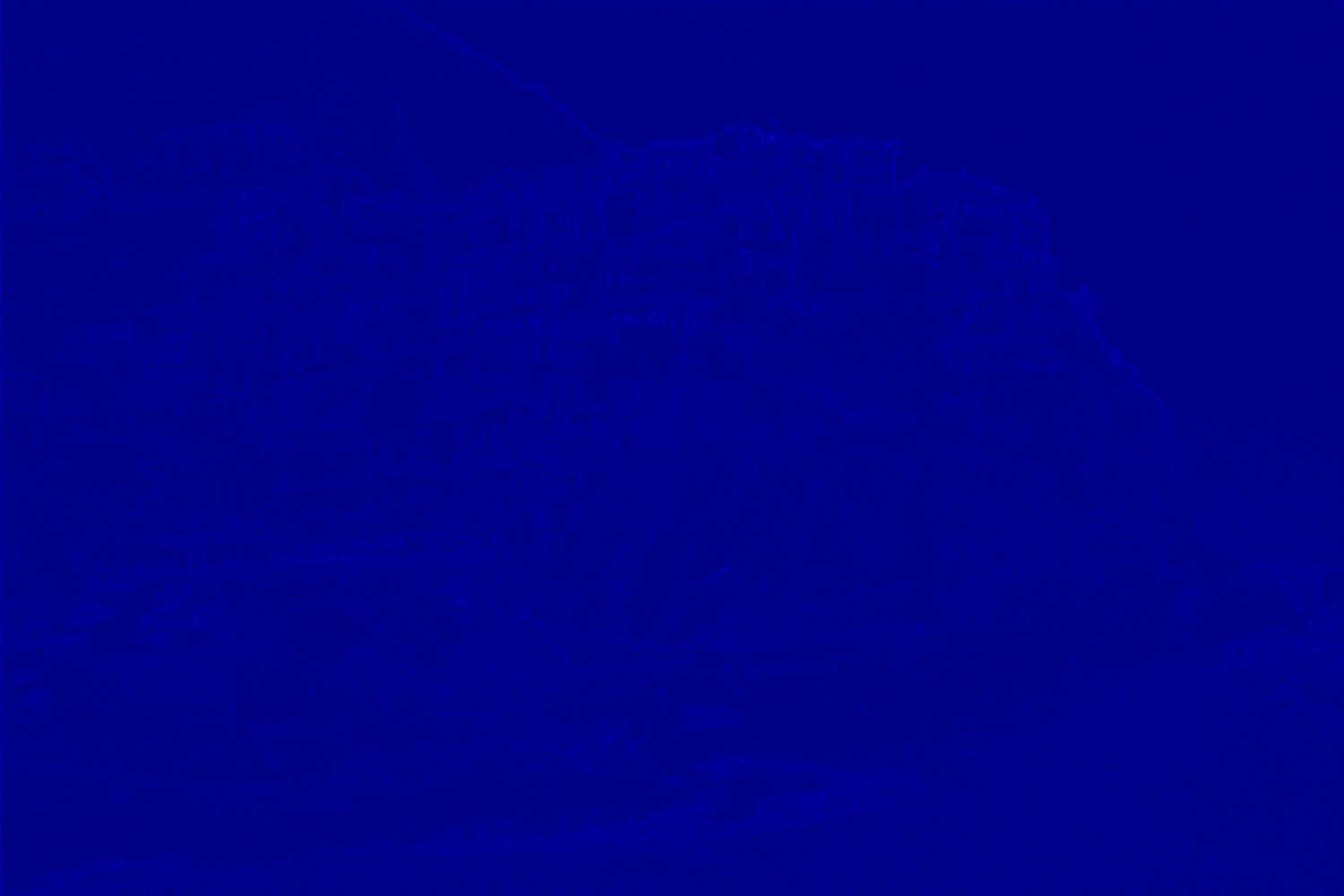}&\\
\rotatebox{90}{\small \hspace{6mm} RGB image}&
\includegraphics[width=0.24\linewidth]{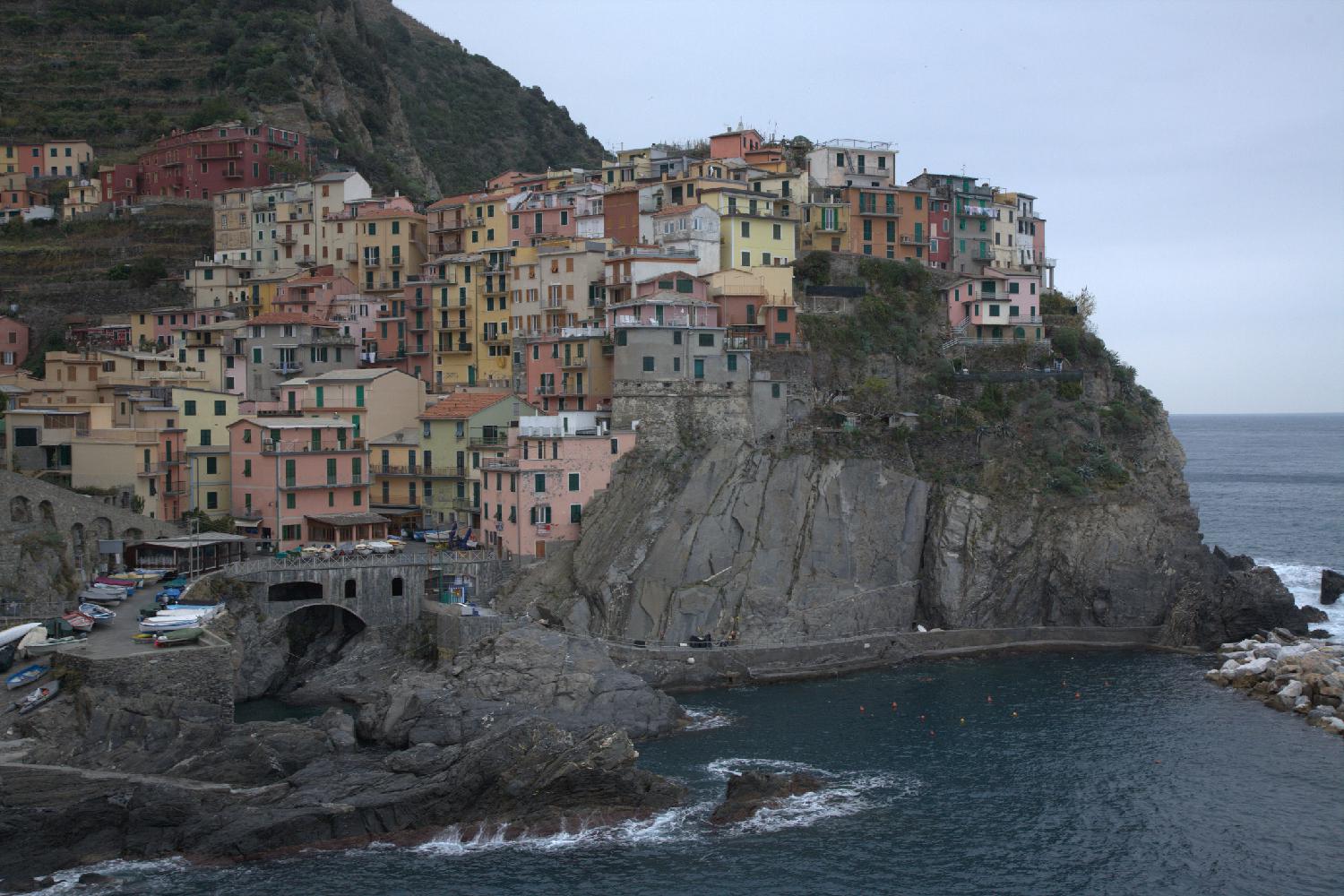}&
\includegraphics[width=0.24\linewidth]{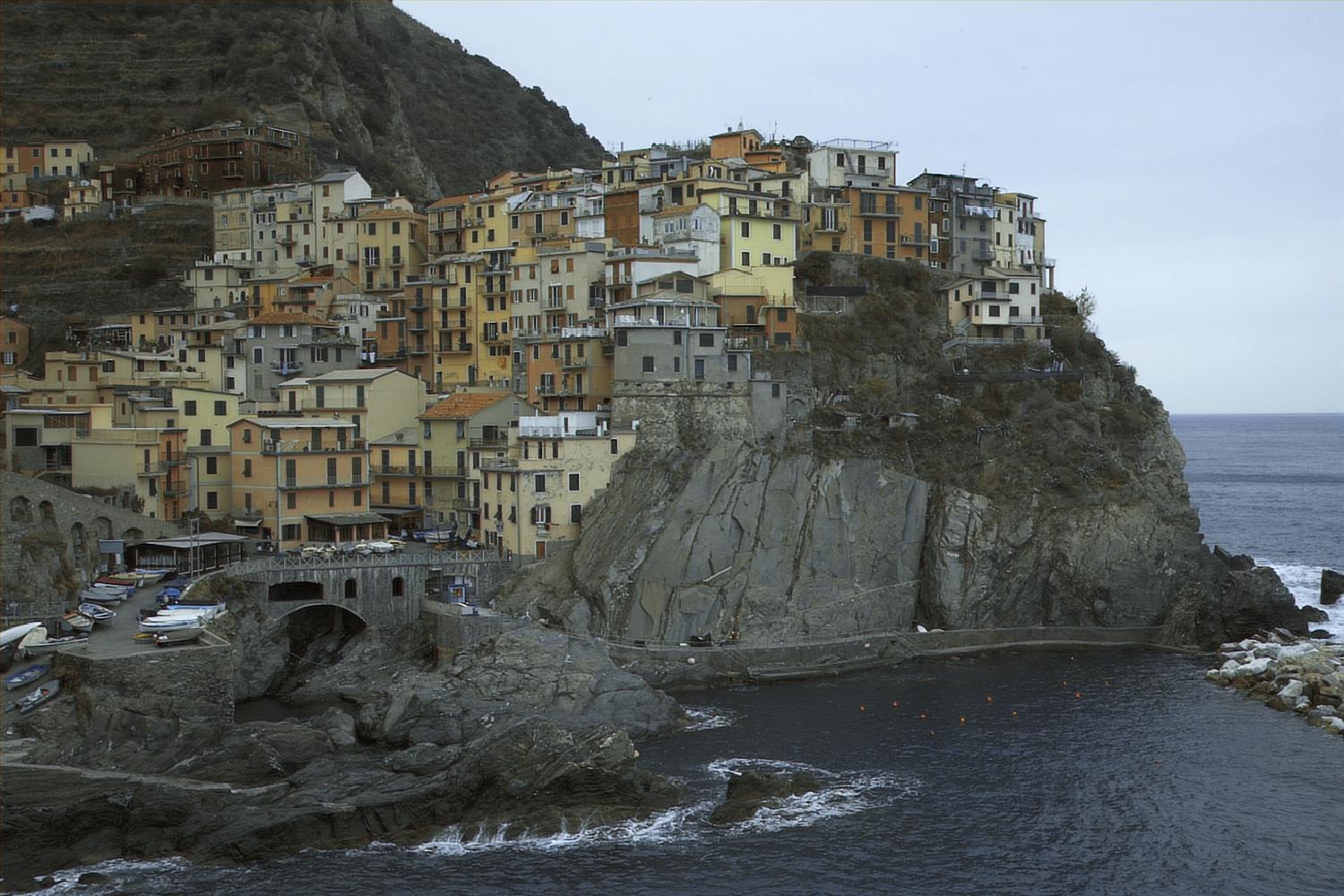}&
\includegraphics[width=0.24\linewidth]{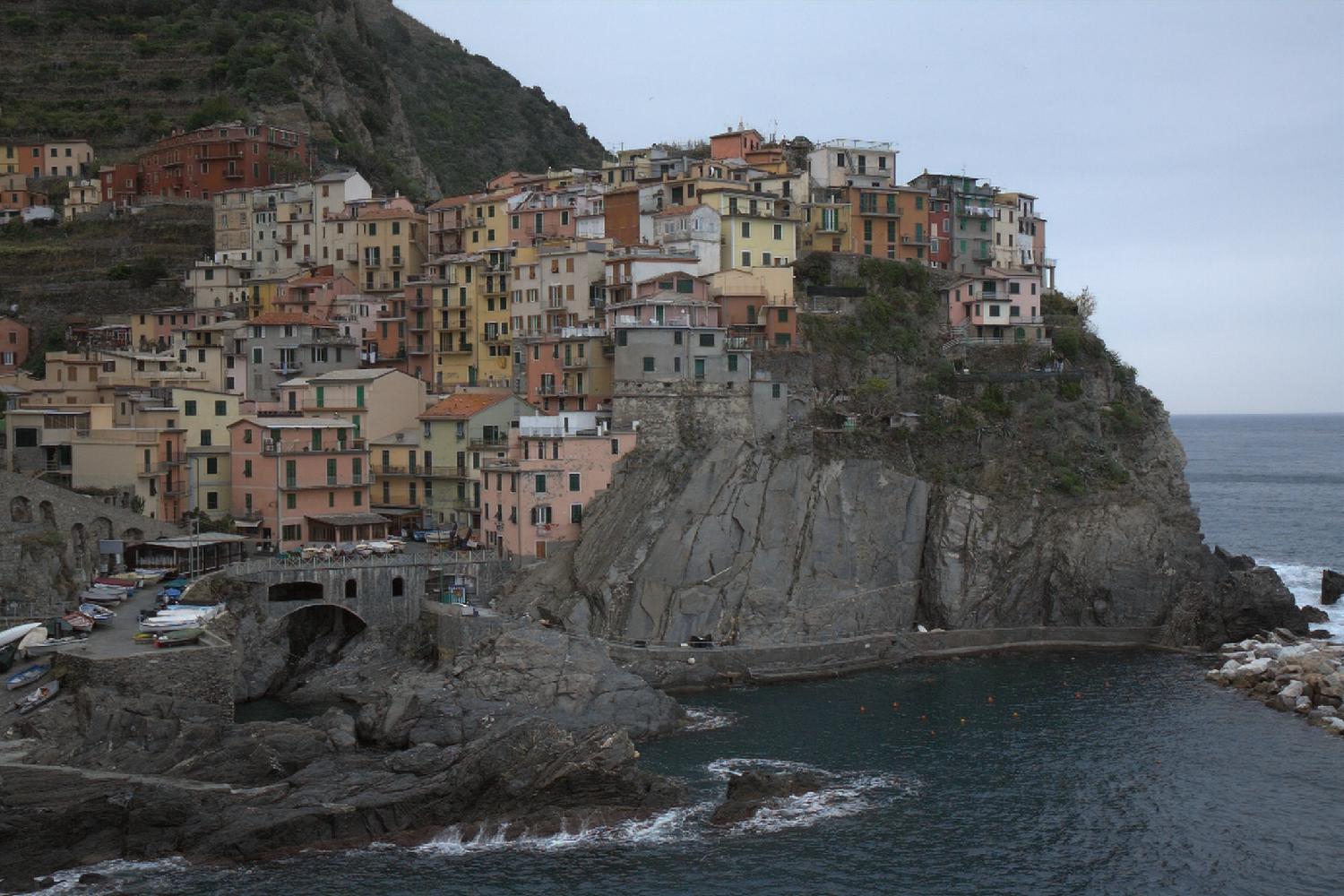}&
\includegraphics[width=0.24\linewidth]{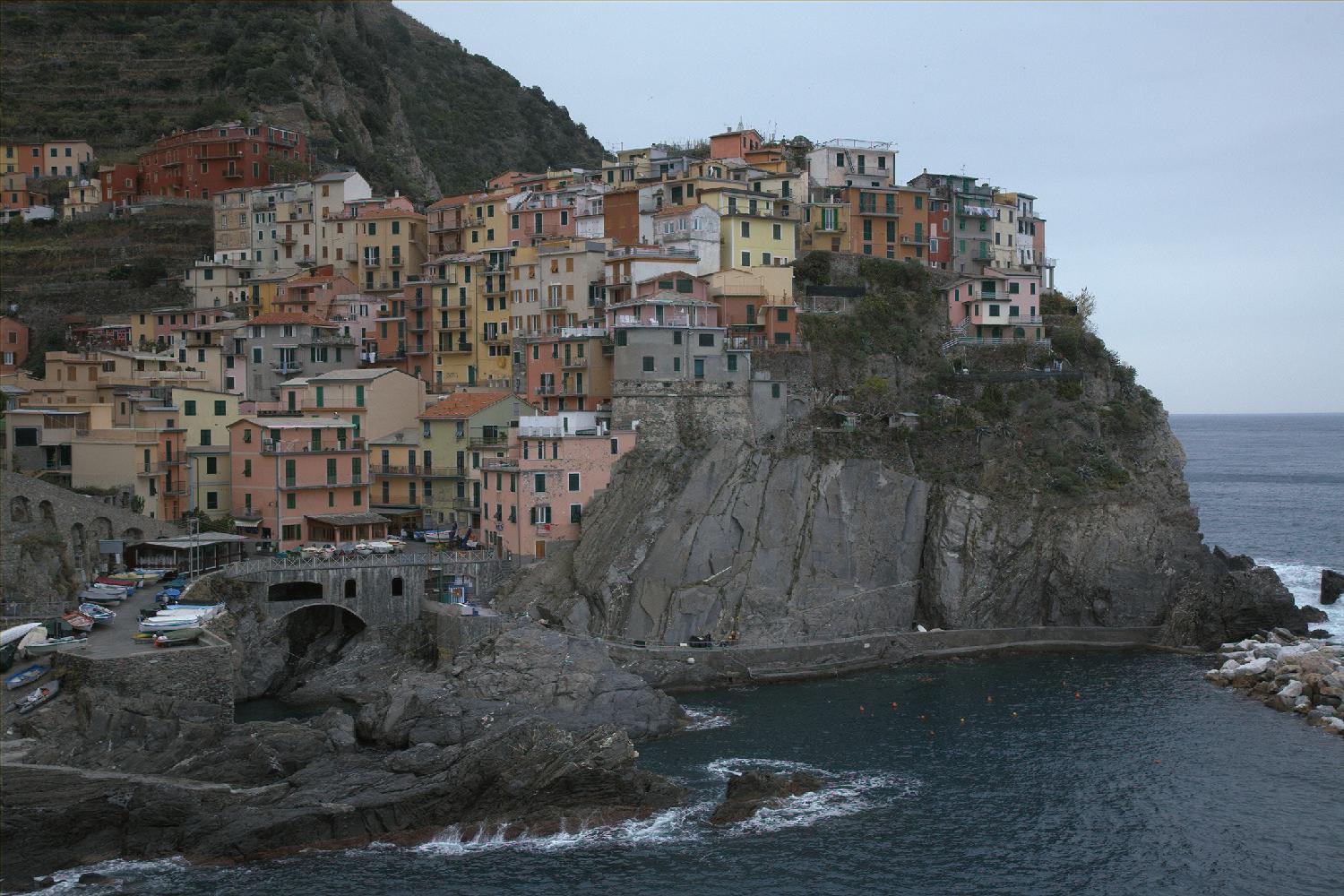}&\\
&Ground truth & Invertible Grayscale~\cite{xia2018invertible} & U-net~\cite{Chen_2018_CVPR} & Ours \\
\end{tabular}
\vspace{1mm}
\caption{Comparison with baselines. Invertible Grayscale~\cite{xia2018invertible} fails at learning a good balance between RGB rendering and RAW recovering, which results in relatively poor performance in both RGB and RAW images. The U-net~\cite{Chen_2018_CVPR} can render comparable RGB performance with ours but perform worse at RAW recovering. Our invertible ISP can both render visually pleasing RGB images and reconstruct realistic RAW data. The GT RAW is visualized through bilinear demosaicing, and other RAW images are visualized through error maps. This figure is best viewed in the electronic version.}
\label{fig:main_comparison}
\vspace{-0.7em}
\end{figure*}

\section{Experiments} 
\subsection{Experimental setup}
\mypara{Datasets.} We collect the Canon EOS 5D subset (777 image pairs) and the Nikon D700 subset (590 image paris) from the MIT-Adobe FiveK dataset~\cite{bychkovsky2011learning} as the training and test data.
We train our model for each camera separately. We randomly split each of the two sets (Canon, Nikon) into training and test sets with a ratio of 85:15. 
We use the LibRaw library to process the RAW images to render ground-truth sRGB images. 
In general, LibRaw conducts most representative ISP steps in modern digital cameras to render sRGB images, including color space conversion, demosaicing, denoising, white balancing, exposure compensation, gamma compression, and global tone mapping.

\mypara{Implementation details.}
We utilize random crop, random rotation, and random flip as data augmentation to train our model. 
We preprocess the raw data using the white balance parameters provided by camera metadata since estimating white balance directly from raw images is a research topic in itself~\cite{barron2015convolutional}. 
To test the effectiveness of our JPEG simulator, we store the ground truth RGB into JPEG format, whose quality is set to 90 (Q=90, most representative JPEG quality in modern digital cameras). We also conduct experiments without preprocessing white balance and with another JPEG quality, whose quantitative results are accessible in the supplement. At test time, we first conduct the forward pass of our network to render RGB images and save them into JPEG images. Then we load the saved JPEG images and conduct the inverse step to recover RAW images. 

\subsection{Baselines} 
\label{sec:baselines}
\mypara{UPI. } Brooks et al.~\cite{brooks2019unprocessing} unprocess the sRGB images to synthesize high-quality RAW images for learned RAW denoising. They adopt camera priors to inverse ISP step-by-step, such as digital gain, tone mapping curves, white balance, and color correction matrices~\cite{brooks2019unprocessing}. Since the metadata such as color correction matrix, white balance, and digital gain are camera dependent, we modified these parameters in their method to fit our dataset. We use their described method to estimate metadata for our datasets.

\mypara{CycleISP.} We select the state-of-the-art learning-based RAW synthesizing method, CycleISP~\cite{zamir2020cycleisp}, as another baseline for synthetic RAW direction. Note that their model has access to RGB images at test time, and thus we only need to compare with their synthesized RAW images. We directly utilize their pretrained model since their framework is trained on the MIT-Adobe FiveK dataset, and their proposed color attention unit can be generalized to different cameras.

\mypara{U-net.} U-net is a representative architecture for ISP in recent year publications~\cite{Chen_2018_CVPR, zhang2019zoom}, thus we implement a encoder-decoder baseline using U-net~\cite{ronneberger2015u}. Both the encoder and decoder are consist of an independent U-net. 
Same as~\cite{Chen_2018_CVPR, zhang2019zoom}, we pack the Bayer pattern RAW into R-G-G-B channels for encoder input and utilize the depth-to-space operation to restore the RGB resolution. We utilize the same data augmentation strategies as our InvISP. We jointly train the encoder and decoder of our U-net baseline using L1 loss from scratch on all our datasets.

\mypara{Invertible Grayscale.} Invertible Grayscale~\cite{xia2018invertible} is a general framework to learn the forward and inverse mapping between two space, such as color-image space and grayscale-image space.
The encoder of Invertible Grayscale takes a 3-channel RGB image as input and processes it to a single-channel grayscale image. The decoder recovers the original sRGB image with the same color from the grayscale image. Similar to their settings, we change the input from sRGB image to RAW data after bilinear demosaicing and set the output of the encoder to the 3-channel RGB image. Since the lightness loss function is not suitable for our tasks, we remove it for our experiments.

\subsection{Results}
\mypara{Quantitative results.}
To quantitatively evaluate our method, we use PSNR and SSIM for rendered RGB images, and PSNR for recovered RAW images. The comparison with baselines is reported in Table~\ref{table:quant_other}. Compared with the RAW synthesizing method UPI and CycleISP, our model can recover more accurate RAW data, which is proved by more than 13 dB improvement of PSNR. The results are not surprising because lots of information lost in the ISP is quite hard to invert, which results in poor performance for synthetic RAW reconstruction methods. However, our InvISP can jointly optimize RGB rendering and RAW recovering process and thus is better to handle the information lost in quantization, JPEG compression, and saturated value clipping problem in ISP. 
For the Invertible Grayscale and the U-net baselines, the results indicate that our method contributes a better ISP as well as a stronger model for recovering RAW data. This is because using two separate networks for ISP and inverse ISP will cause the error accumulation problem, which further degrades the RAW reconstruction performance. Our methods take the inherent reversibility of invertible neural networks thus can recover higher-quality RAW images than baselines.

\begin{figure}
\centering
\begin{tabular}{@{}c@{\hspace{0.7mm}}c@{\hspace{0.5mm}}c@{\hspace{0.2mm}}c@{\hspace{0.2mm}}c@{}}
\rotatebox{90}{\small \hspace{5mm} Retouching}& 
\includegraphics[width=0.475\linewidth]{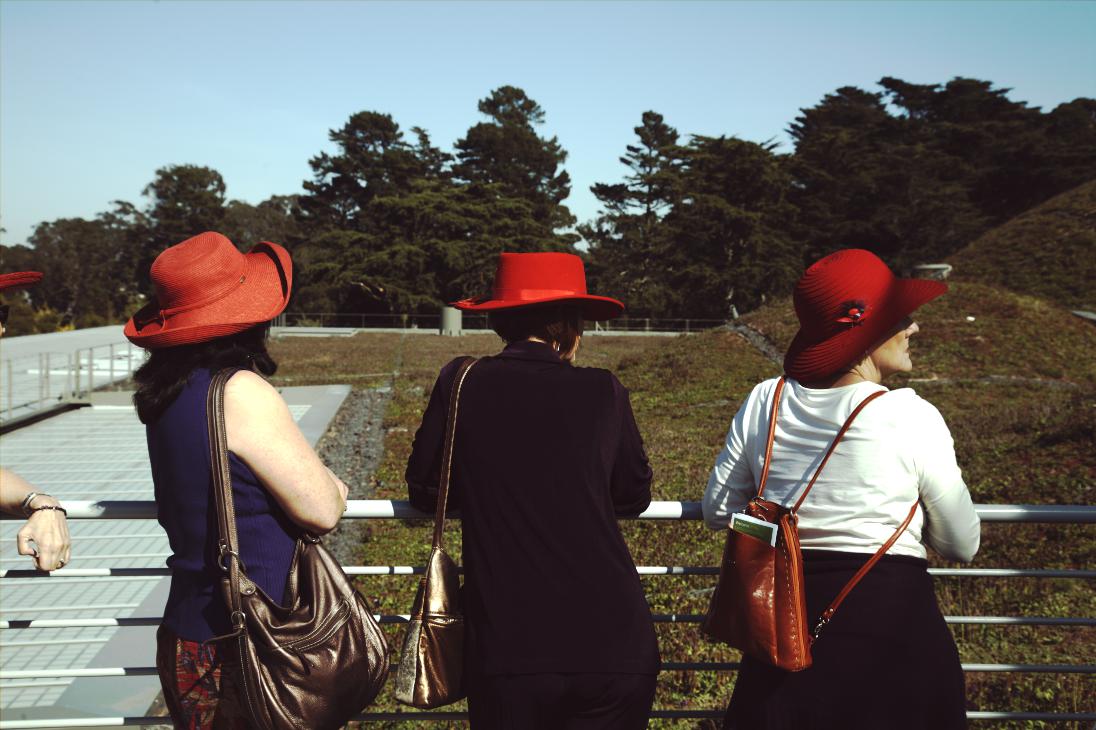}&
\includegraphics[width=0.475\linewidth]{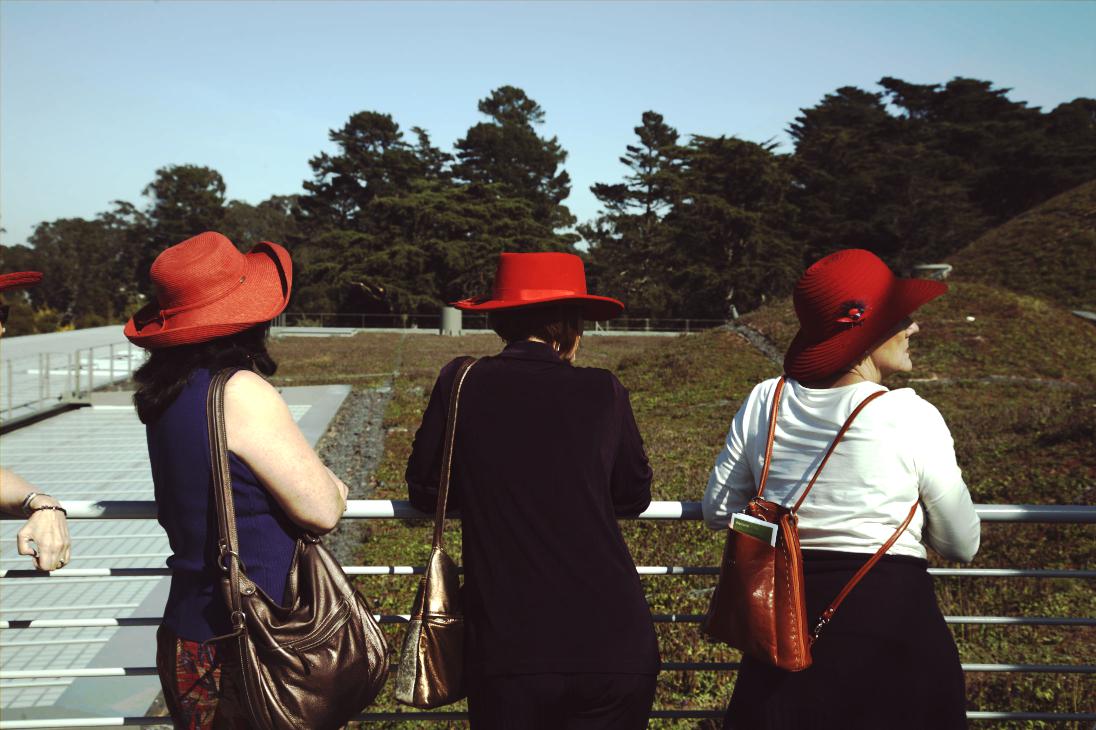}&\\
\rotatebox{90}{\small \hspace{-1mm} HDR reconstruction}& 
\includegraphics[width=0.475\linewidth]{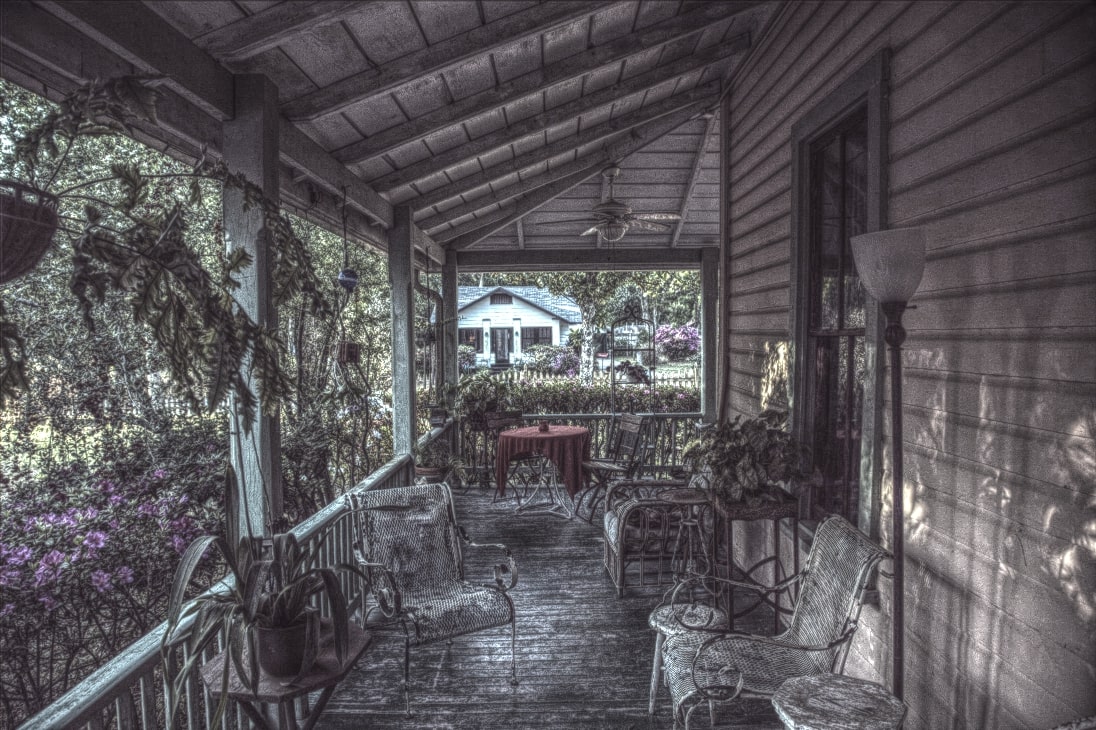}&
\includegraphics[width=0.475\linewidth]{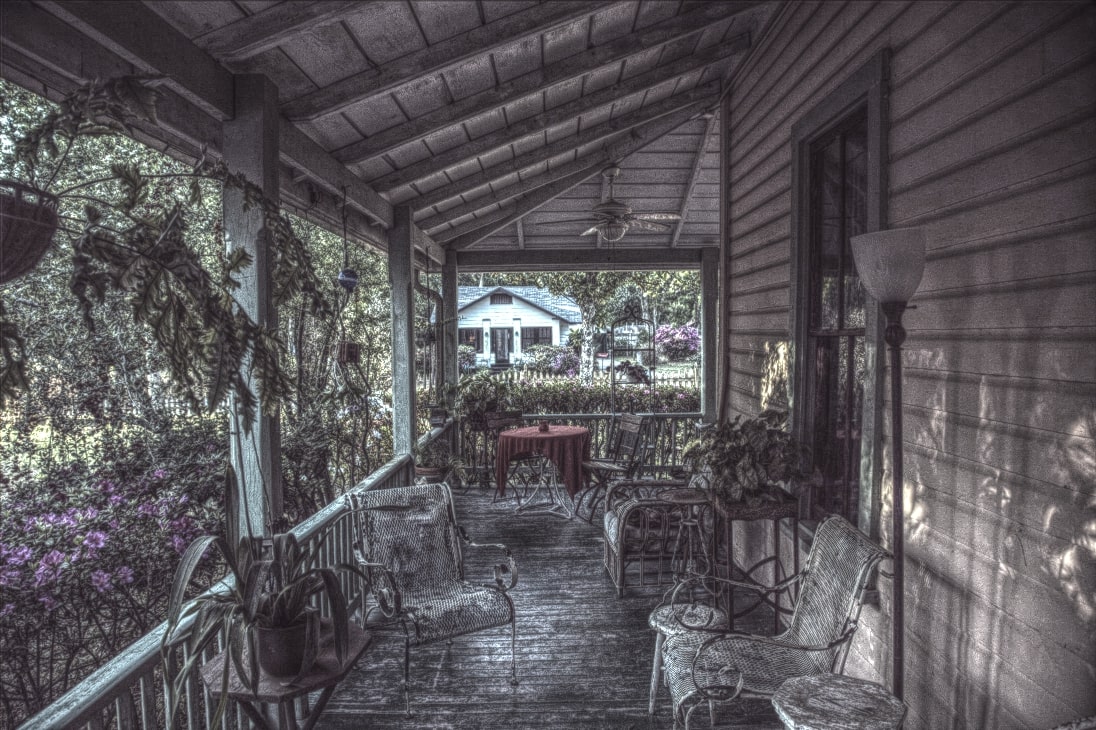}&\\
&Ground truth & Our result \\
\end{tabular}

\vspace{1mm}
\caption{Our model can enable image retouching~\cite{hu2018exposure} and HDR reconstruction~\cite{paris2011local} applications. Note that we use the pretrained model of~\cite{hu2018exposure} for retouching.}
\label{fig:exposure}
\vspace{-0.7em}
\end{figure}

\mypara{Qualitative results.} 
We show qualitative comparisons against baseline methods in Figure~\ref{fig:cycleisp} and Figure~\ref{fig:main_comparison}. 
In Figure~\ref{fig:cycleisp}, the synthetic RAW by CycleISP and UPI differs a lot from ground truth RAW images, especially at over-exposed regions, which indicates that their model performs poorly to handle the information loss of ISP. Our model, however, can recover the RAW information much better than synthetic RAW methods, even at challenging highlight pixels, which raises the potential for prospective photo editing tasks. 
In Figure~\ref{fig:main_comparison}, Invertible Grayscale fails to pursue a good balance between RGB rendering and RAW reconstruction. Our naive U-net baseline can achieve comparable performance in terms of RGB rendering but not perform well at RAW recovering. Our method reconstructs higher-quality RAW images on edges and over-exposed areas without sacrificing the RGB rendering performance.

\section{Applications} 
\subsection{RAW data compression}
One important application of our framework is RAW data compression for cameras. Traditionally, users need to explicitly store RAW data for further applications. Using our technique, however, only JPEG images need to be stored, and users can reconstruct the corresponding RAW data from JPEG images. 
To evaluate the reduced file size, we calculate the compression ratio and the bit per pixel (BPP). The compression ratio $C_{ratio}$~\cite{poynton2012digital} is calculated by
\begin{align}
 C_{ratio} = \frac{\text{Uncompressed size}}{\text{Compressed size}} = \frac{B_{BMP}}{B_{JPEG}}   ,
\end{align}
where $B_{BMP}$ is the file size of RAW data in BMP format and $B_{JPEG}$ is the file size of rendered sRGB image in JPEG format. Note that $B_{BMP}$ is calculated by~\cite{bourke1998bmp}
\begin{align}
B_{BMP} = 54 + \frac{H \times W \times b}{8}  ,
\end{align}
where $H$, $W$ and $b$ are the height, width and the bit depth of the RAW data. 
We further compare our compression effectiveness with Adobe lossy DNG. As shown in Table ~\ref{table:compression}, the file size is highly reduced, even compared with lossy DNG. 


\begin{table}[t!]
\centering
\setlength{\tabcolsep}{3mm}
\ra{1.25}
\begin{tabular}{@{}l@{\hspace{2mm}}c@{\hspace{2mm}}c@{\hspace{3mm}}c@{\hspace{1mm}}c@{\hspace{1mm}}c@{\hspace{3mm}}c@{\hspace{3mm}}c@{\hspace{2mm}}c@{\hspace{2mm}}c@{1mm}}
\toprule
 & \multicolumn{2}{@{\hspace{-1mm}}c}{Compression ratio $\uparrow$} & \multicolumn{2}{@{\hspace{-1mm}}c}{BPP $\downarrow$} \\
 Dataset & Lossy DNG & Ours & Lossy DNG & Ours\\
\midrule 
NIKON D700 & 1.61 &34.98 &8.73 & 0.4655 \\
Canon EOS 5D &  1.52&27.37 & 6.56&0.5237 \\
\bottomrule 
\end{tabular}
\vspace{1mm}
\caption{Comparison of the compression ratio of the file size and bit per pixel (BPP) between our method and lossy DNG. The file size is significantly reduced by our framework.}
\label{table:compression} 
\vspace{-0.7em}
\end{table} 


\subsection{Image retouching}
Professional photographers choose to retouch images from RAW data for better visual quality. 
We demonstrate that our recovered RAW can be directly taken as input for high-quality image retouching. We use an automatic deep learning based image retouching method Exposure~\cite{hu2018exposure} as an example. We preprocess the recovered RAW and ground truth RAW through demosaicing and white balancing, following the setting of the paper~\cite{hu2018exposure}.
We directly utilize their pretrained model that is also trained on the MIT-Adobe FiveK dataset. As illustrated in Figure ~\ref{fig:exposure}, our reconstructed RAW data has an indistinguishable visual quality to the RAW data captured by the camera.

\subsection{HDR reconstruction and tone manipulation}
Inferring a high dynamic range image from a single low dynamic range input is challenging~\cite{eilertsen2017hdr, endo2017deepReverseTone} since the information lost in saturated and under-exposed regions are hard to invert accurately. Our invertible ISP framework fundamentally alleviates these difficulties and thus enables single image HDR reconstruction. Further, the recovered HDR image can be tone mapped to display much more details than the original RGB input. In Figure~\ref{fig:exposure}, we use~\cite{paris2011local} as tone mapper to demonstrate the potential of our method.

\section{Conclusion} 
We have proposed an end-to-end invertible image signal processing (InvISP) framework to generate visually pleasing RGB images and recover nearly perfect quality RAW data. 
We leverage the idea from invertible neural networks to design our invertible structure and integrate a differentiable JPEG simulator to enhance the network stability to JPEG compression. We use LibRaw to simulate ground-truth ISP on the MIT-Adobe FiveK dataset. 
We evaluate our method through comparisons with other frameworks and RAW data synthesis methods. 
We also demonstrate that our framework enables RAW data compression, image retouching, and HDR reconstruction tasks. 
We hope our method can inspire further research on RAW image reconstruction. 


{\small
\bibliographystyle{ieee_fullname}
\bibliography{egbib}
}

\end{document}

%% file: vcl-shortcuts.tex
\usepackage{times}
\usepackage{epsfig}
\usepackage{graphicx}
\usepackage{float}
\usepackage{wrapfig}
\usepackage{amsmath,amssymb,amsthm}
\usepackage{algorithm,algorithmicx,algpseudocode}
\usepackage{bm,xspace}
\usepackage{comment}
\usepackage{verbatim}
\usepackage{multirow}
\usepackage{balance}
\usepackage{url}
\usepackage{booktabs}
\usepackage{etoolbox,siunitx}
\usepackage{calc}
\usepackage{pifont,hologo}
\usepackage[usenames, dvipsnames]{color}
\usepackage{nicefrac}

\newcommand{\ra}[1]{\renewcommand{\arraystretch}{#1}}

\usepackage[super]{nth}
\usepackage{nicefrac}
\robustify\bfseries

\DeclareMathSymbol{@}{\mathord}{letters}{"3B}

\newcommand\mypara[1]{\vspace{1mm}\noindent\textbf{#1}}

\def\latex/{\LaTeX}
\def\bibtex/{\hologo{BibTeX}}
